\documentclass[review]{elsarticle}

\usepackage{lineno,hyperref}
\usepackage{listings}
\usepackage{todonotes}
\usepackage{adjustbox}
\usepackage{tabularx}
\usepackage{comment}
\usepackage{enumitem}
\usepackage{graphicx}
\usepackage{subcaption}
\usepackage{tabularx}
\usepackage{scrextend}
\usepackage{float}
\modulolinenumbers[5]

\lstdefinestyle{customJava}{
    language={Java},
    frame=none,
    basicstyle=\footnotesize\ttfamily, %
    identifierstyle=\color{blue},
    keywordstyle=\bfseries\color{green!40!black},
    commentstyle=\itshape\color{purple!40!black},
    columns=flexible, %
    extendedchars=true, %
    showspaces=false, %
    showstringspaces=false, %
    linewidth=\linewidth,
    tabsize=2,
    numbers=left,
    xleftmargin=7mm,
    framexleftmargin=6mm,
    literate=%
         {ç}{{\c{c}}}1
         {á}{{\'a}}1
         {ã}{{\~a}}1
         {Ã}{{\~A}}1
         {í}{{\'i}}1
         {é}{{\'e}}1
         {ú}{{\'u}}1
         {ó}{{\'o}}1
         {Á}{{\'A}}1
         {Í}{{\'I}}1
         {É}{{\'E}}1
         {Ú}{{\'U}}1
         {Ó}{{\'O}}1
}
\journal{Information and Software Technology}

\begin{document}
\newcommand{\myparagraph}[1]{\vspace{.2em}\noindent\textit{\textit{\textbf{#1}}}\hspace{.3em}}

\begin{frontmatter}

\title{CADV: A software visualization approach for code annotations distribution}
\tnotetext[mytitlenote]{Fully documented templates are available in the elsarticle package on \href{http://www.ctan.org/tex-archive/macros/latex/contrib/elsarticle}{CTAN}.}

\author[unifei,inpe]{Phyllipe Lima}
\ead{phyllipe@unifei.edu.br}

\author[unibz]{Jorge Melegati}
\ead{jorge@jmelegati.com}

\author[ufabc]{Everaldo Gomes}
\ead{everaldogjr@gmail.com}

\author[inatel]{Nathalya Stefhany Pereira}
\ead{nathalya.stefhany@gec.inatel.br}

\author[unibz]{Eduardo Guerra}
\ead{eduardo.guerra@unibz.it}

\author[ufabc]{Paulo Meirelles}
\ead{paulo.meirelles@ufabc.edu.br}

\address[unifei]{Federal University of Itajubá - IMC - UNIFEI}
\address[inpe]{National Institute for Space Research -- INPE}
\address[inatel]{National Institute for Telecommunications -- Inatel}
\address[unibz]{Free University of Bolzano-Bolzen -- UniBZ}
\address[ufabc]{Federal University of ABC -- CMCC-UFABC}

\begin{abstract}
\textit{Context}: Code annotations is a widely used feature in Java systems to configure custom metadata on programming elements. Their increasing presence creates the need for approaches to assess and comprehend their usage and distribution. In this context, software visualization has been studied and researched to improve program comprehension in different aspects.\\
\textit{Objectives}: This study aimed at designing a software visualization approach that graphically displays how code annotations are distributed and organized in a software system and developing a tool, as a reference implementation of the \textcolor{black}{approach}, to generate views and interact with users. \\ 
\textit{Methods}: We conducted an empirical evaluation through questionnaires and interviews to evaluate our visualization approach considering four aspects: (i) effectiveness for program comprehension, (ii) perceived usefulness, (iv) perceived ease of use, and (iv) suitability for the intended audience. The resulting data was used to perform a qualitative and quantitative analysis.\\ \textit{Results}: The tool identifies package responsibilities providing visual information about their annotations at different levels. Using the developed tool, the participants achieved a high correctness rate in the program comprehension tasks and performed very well in questions about the overview of the system under analysis. Finally, participants perceived that the tool is suitable to visualize the distribution of code annotations. \\ 
\textit{Conclusions}: The results show that the visualization approach using the developed tool is effective in program comprehension tasks related to code annotations, which can also be used to identify responsibilities in the application packages. Moreover, it was evaluated as suitable for newcomers to overview the usage of annotations in the system and for architects to perform a deep analysis that can potentially detect misplaced annotations and abnormal growths on their usage. 
\end{abstract}


\end{frontmatter}


\section{Introduction}

Code annotations were introduced in version 5 of Java to configure custom metadata directly on programming elements, such as methods and classes. Tools or frameworks usually consume this metadata to gather additional information about the software, allowing the execution of more specific behavior. Since annotations are inserted directly into the source code, they are a convenient and quick alternative to configure metadata. Core enterprise Java APIs make extensive use of code annotations, making them a relevant feature used daily by developers. 

A recent study performed in Java open source projects~\cite{Lima2018jss} identified at least one code annotation in 78\% of the classes. Another study demonstrated that code annotations are actively maintained, and most code annotations changes are consistent with other code changes~\cite{Yu2019}. This evidence suggests that code annotations are a feature that has a non-neglectable impact on software maintenance. 

According to Rajlich~\cite{rajlich2014}, software evolution and maintenance is the most costly and challenging activity in the software development life cycle. While the comprehension of a single module depends directly on the code quality, it is not straightforward to understand one aspect across the whole source code. 
Given that developers spend most of their time comprehending the software to be able to add and maintain features~\cite{Hasselbring2020}, the increasing presence of code annotations creates the need for approaches that can help in understanding how this feature is used through the system components. 

The approach of software visualization has become increasingly used to support developers in software comprehension~\cite{francese2016,merino2018}. Some approaches aim to represent software as a known environment, such as a city~\cite{Wettel2011,romano2019} or a forest~\cite{erra2012}. Another technique is to create what is known as a polymetric view, defined as a lightweight visualization enriched with software metrics~\cite{lanza2003}. 

To aid the software comprehension of code annotations usage in software systems, we propose a visualization approach named Code Annotations Distribution Visualization (CADV). This visualization uses code annotation metrics from our previous work~\cite{Lima2018jss} as input and the circle packing approach~\cite{Kenneth2007} to draw the system under analysis. The CADV is composed by three different views: \textbf{System View}, \textbf{Package View}, and \textbf{Class View}. They complement each other and display code annotations information within different scopes and granularity. The \textbf{System View} displays annotations distributed by packages, the \textbf{Package View} displays annotations distributed in the classes of a single package, and the \textbf{Class View} displays the distribution of annotations in the code elements within a class. The views use colors to represent the annotation schemas and circle size to represent metric values. As a reference implementation of the CADV, we developed an open-source web tool called Annotation Visualizer (AVisualizer)\footnote{\url{https://github.com/metaisbeta/avisualizer}}.

We conducted an empirical evaluation through questionnaires with students and interviews with professional developers to assess CADV using AVisualizer. We built the survey to evaluate its effectiveness in program comprehension tasks, its suitability for the intended target audience of the tool, and how useful and easy the respondents felt about the approach. Combining the results from the questionnaire and interview, we conducted a qualitative and quantitative analysis to reach our conclusions about the usage of the proposed approach in the mentioned aspects.

Based on these studies, we observed a high correctness rate in questions related to a general view of the system. Furthermore, participants could visualize package responsibilities by associating them with annotations usage. The strategy applied to display code annotation also allowed participants to detect potentially misplaced annotations. \textcolor{black}{The interview with developers also suggests that the AVisualizer may outperform code inspection using standard IDE code search tools in code comprehension tasks related to annotations.} Finally, the circle packing approach was shown suitable to display the hierarchical structure of the analyzed source code.

\textcolor{black}{The remaining of this paper is organized as follows: Section~\ref{sec:background} presents our background and motivations for this work; Section~\ref{sec:related-work} presents related works that explored software visualization to improve software comprehension; Section~\ref{sec:rd-visu} describe the research design used to propose and evaluate the CADV;  Section~\ref{sec:cadv} defines the three views from the CADV approach and their implementation in the AVisualizer tool; Section~\ref{sec:evaluation} describes the design of the study conducted to evaluate the CADV; Section~\ref{sec:discussion} presents the study results and discussions about them; finally, in Section~\ref{sec:conclusion}, we conclude the paper, highlighting the contributions and pointing some directions for future studies.} 
\section{Background and Motivation}
\label{sec:background}

\textcolor{black}{In this section, we introduce code annotations and how they are used to configure custom metadata in programming elements (Section~\ref{ssec:annotations}). Then, in Section~\ref{ssec:annotmetric}, we present the suite of software metrics for code annotations proposed and defined in \cite{Lima2018jss}. These metrics aim to measure the characteristics of code annotations we use as input for the software visualization approach (CADV) proposed in this paper. In the final section, we present other works that have assessed code annotations and how their results reinforce the relevance and impact of code annotations, motivating our work of developing an approach to visualize code annotations and their distribution.}

\subsection{Code Annotations and Metadata Configuration}
\label{ssec:annotations}

The term ``metadata'' is used in various contexts in Computer Science. In all of them, it means data referring to the data itself. When discussing databases, while ``data" refers to the domain information persisted, ``metadata" refers to their description, i.e., the table's structure. In the object-oriented context, the data are the instances, and the metadata is their description, i.e., information describing the class. As such, fields, methods, super-classes, and interfaces are all metadata of a class instance. A class field, in turn, has its type, access modifiers, and name as its metadata. When a developer uses reflection, it is manipulating the metadata of a program and using it to work with previously unknown classes~\cite{guerra2010}.

Some programming languages provide features that allow custom metadata to be defined and included directly on programming elements. This feature is supported in languages like Java and C\#, which are referred to as code annotations and attributes. They were integrated into the Java language in version 1.5, becoming popular in the development community. Some well-known and established APIs and frameworks such as EJB (Enterprise Java Beans), JPA (Java Persistence API), JUnit, and Spring Boot use annotations extensively. This native annotation support encourages many Java frameworks and API developers to adopt the metadata-based approach in their solutions. 

\textcolor{black}{Two recent studies that measured their occurrence in the source code highlight the importance of studying code annotations. One of them evaluated the top-ranked Java projects hosted on GitHub and observed that, on average, 76\% of classes have at least one annotation~\cite{Lima2018jss}. The other study included 1,094 projects in the analysis and found that the median value of annotations per project is 1,707~\cite{Yu2019}. Both studies detected possible abuses in the usage of annotations, finding a high number of annotations in individual classes and projects. This evidence shows that annotations are a popular feature in Java projects and consequently can influence several aspects of the code and software projects as a whole.} 

When discussing annotation-based APIs in the context of metadata-based frameworks, such as JPA, JUnit, and Spring, an important concept is ``annotation schema''\footnote{in this work we also refer to it simply as ``schema''}. Since one annotation rarely is enough to represent the information needed in a metadata-based API~\cite{guerra2016annotationapis}, usually a group of annotations is combined in the same package to fulfill that requirement. So, ``annotation schema'' can be defined as the group of annotations used to represent the metadata structure used by a metadata-based API.

Since these annotations, which are part of the same metadata structure, are often part of the same package, this information is used as a heuristic to identify an annotation schema~\cite{Lima2018jss}. In other words, annotations from the same package are considered part of the same schema. Consequently, a single metadata-based framework may contain several schemas, usually representing distinct data structures responsible for representing metadata for different framework features. 

\textcolor{black}{To better illustrate the definition of ``schema'', consider the code in Figure~\ref{fig:anot-example-schema} that presents a simple class responsible for executing unit tests using the JUnit 4 framework. The annotations \texttt{@Test}, \texttt{@After}, \texttt{@Before} that belong to the package \texttt{org.junit} are used in combination to define how the framework should execute the tests, being part of the same annotation-based API. In this case, we refer to this annotation set as part of the \texttt{org.junit} annotation schema. In practical terms, an automated approach for extracting the schemas used in classes can identify each package from the imported annotations as a separate schema~\cite{Lima2020}.}

\lstset{style=customJava} 
\begin{figure}[ht]

\begin{minipage}{\linewidth}
\begin{lstlisting}
import org.junit.After;
import org.junit.Before;
import org.junit.Test;

public class TestClass {
    @Before
    public void setUp(){
        //initializations
    }
    @Test
    public void testMethod(){
        //Execute tests
    }
    @After
    public void cleanTest(){
        //clear resources allocated during initialiazation
    }
}
\end{lstlisting}
\end{minipage}
\caption{Example with \textit{org.junit} schema.}
\label{fig:anot-example-schema}
\end{figure}

The code example in Figure~\ref{fig:anot-example-schema} makes clear the strong relationship between the presence of annotations of a given schema and code responsibilities. It is clear that a class with annotations from the \texttt{org.junit} schema is responsible for testing. 
\textcolor{black}{Therefore, detecting their presence in software systems may help understand how developers organize packages' responsibilities. As another example, if the schema \texttt{javax.persistence} is imported in a class, we can infer that this class is performing actions related to object-relational mapping and dealing with a database. We also would expect that these classes would be grouped in the same package, which may help identify the responsibility of the package instead of a single class, giving a broader understanding of the system under analysis. In other words, the presence of classes with annotations of a given schema in a package might indicate their role in the system architecture. Some studies confirmed this possibility by performing Java source code static analysis using code annotations to identify the classes' architectural role~\cite{aniche-emse-mvc-smells, aniche-satt-code-metric-thresholds}. Finally, this information also helps understand the coupling between applications to specific annotation-based APIs or metadata-based frameworks. In the first example (org.junit) the class is coupled to the JUnit framework, and in the second example (javax.persistence), the class is coupled to the JPA.}

\textcolor{black}{The software visualization that we are proposing and discussing in Section~\ref{sec:cadv} was highly motivated by this relationship between annotation schema and class/package responsibilities. The visualization aims to display the used schemas so developers may visualize the responsibilities of classes and packages in the target system.}

\subsection{Measuring Code Annotations}
\label{ssec:annotmetric}

The visualization approach we propose - Code Annotations Distribution Visualization (CADV) - uses source code metrics values as input. Source code metrics help summarize particular aspects of software elements, detecting outliers in large amounts of code. They are valuable in software engineering since they aid in monitoring code quality and controlling complexity~\cite{Lanza2006}, making developers aware of possible abnormal growth of specific system characteristics.

The metrics used in CADV are a suite dedicated to measuring code annotations proposed in our previous work~\cite{Lima2018jss}. This subsection describes these metrics briefly to clarify how they are used in our visualization. The description focuses only on the metrics used in CADV. The code presented in Figure~\ref{fig:ac-metrics} is used as an example of how the metrics' values would be collected. 

\begin{figure}

\begin{minipage}{\linewidth}
\begin{lstlisting}
import javax.persistence.AssociationOverrides;
import javax.persistence.AssociationOverride;
import javax.persistence.JoinColumn;
import javax.persistence.NamedQuery;
import javax.persistence.DiscriminatorColumn;
import javax.ejb.Stateless;
import javax.ejb.TransactionAttribute;

@AssociationOverrides(value = {
      @AssociationOverride(name="ex",
         joinColumns = @JoinColumn(name="EX_ID")),
      @AssociationOverride(name="other",
         joinColumns = @JoinColumn(name="O_ID"))})
@NamedQuery(name="findByName",
      query="SELECT c " +
            "FROM Country c " + 
            "WHERE c.name = :name")
@Stateless
public class Example {...

   @TransactionAttribute(SUPPORTS)
   @DiscriminatorColumn(name = "type", discriminatorType = STRING)
   public String exampleMethodA(){...}

   @TransactionAttribute(SUPPORTS)
   public String exampleMethodB(){...}

}
\end{lstlisting}
\end{minipage}
\caption{Code to exemplify annotation metrics extracted from \cite{Lima2018jss}}
\label{fig:ac-metrics}
\end{figure}

\begin{itemize}
    \item Annotations in Class (AC): It represents the number of annotations declared on all code elements in a class, including nested annotations. In our code example, the value of AC is equal to 10.
    
    \item Annotations Schemas in Class (ASC): It counts the number of different schemas present in a class, measuring how coupled a class is to different frameworks. This value is obtained by tracking the imports used for the annotations. In the code example, the ASC value is 2. The import \texttt{javax.persistence} is a schema provided by the JPA API, and the import \texttt{javax.ejb} is provided by the EJB API.
    
    \item Arguments in Annotations (AA): It counts the number of arguments contained in an annotation. Each annotation has its own value for the AA metric. For instance, the \texttt{@AssociationOverrides} has only one argument named \texttt{value}, so AA has the value of 1. As another example, \texttt{@AssociationOverride}, contains two arguments, \texttt{name} and \texttt{joinColumns}, so the value for the AA metric is 2.
    
    \item Annotations in Element Declaration (AED): It counts how many annotations are declared in a code element, including nested annotations. In the code example, the method \texttt{exampleMethodA} has an AED value of 2, as it is annotated by \texttt{@TransactionAttribute} and \texttt{@DiscriminatorColumn}.
    
    \item LOC in Annotation Declaration (LOCAD): LOC (Lines of Code) is a well-known metric that counts the number of code lines in a given context. This metric is a variant of LOC that counts the number of lines used in an annotation declaration. As examples, \texttt{@AssociationOverrides} has a LOCAD value of 5, while \texttt{@NamedQuery} has LOCAD equals 4.
    
\end{itemize}

With this suite of metrics, it is possible to characterize the usage of code annotations in a given system, capturing their size and complexity and the coupling generated by them. In our software visualization approach, these values will be used to render the system. Therefore, these metrics can be considered the basis for the CADV approach.

\subsection{Code Annotations Assessment}
\label{rw-annotationassessment}

\textcolor{black}{In our previous work~\cite{Lima2018jss}, we defined a novel suite of software metrics (as described in Section~\ref{ssec:annotmetric}). This work also describes an exploratory data analysis that aimed to understand how these metrics values were distributed in open-source projects by collecting the metrics values from 25 selected projects. It was identified that 90\% of the classes use up to 11 annotations, often declared using one line of code and one argument. Finally, they are usually not nested, being rare to find a two-level nesting. However, some outlier values found in this study indicated potential problems that at least deserve a closer inspection. For instance, we found classes with more than 700 annotations and one annotation taking 58 lines of code. Also, we found that one single code element had 27 annotations configuring it. This result is evidence of extremely high values for annotation metrics that, like other instances of large size metrics values as an extensive method or a high number of parameters~\cite{Fowler1999}, might negatively influence code maintenance.} 

\textcolor{black}{Yu et al.~\cite{Yu2019} performed a large-scale empirical study on code annotations usage, evolution, and impact. The authors collected data from 1094 open-source Java projects and conducted a historical analysis to assess changes in code annotations. The study found huge values for the number of annotations considering the code element, the proportion in terms of lines of code, and the project as a whole. For instance, the authors detected that some code elements might have 41 annotations configuring them, which they considered extremely high. These results indicate the existence of abuses of this feature and confirm the result from Lima et al.~\cite{Lima2018jss}. Another finding revealed that code annotations are constantly changed during project evolution, implicating that some developers use annotations subjectively and arbitrarily, introducing code problems. They also demonstrated that code annotations are deleted because they are redundant (15.4\%) or incorrectly placed (24.5\%).} 

Concerning annotation repetition, Teixeira et al.~\cite{teixeira2018does} performed a study that investigated the source code of a web application by searching for annotations repeated throughout the source code. The findings revealed that some annotations were repeated around 100 times in code elements with shared characteristics in the target system. The repetition of annotations is also mentioned in Yu et al.~\cite{Yu2019}, but no quantitative evidence was presented. The study suggested that more general definitions, such as application-specific code conventions, could significantly reduce the number of configurations.

Based on these studies, it is possible to state that code annotations are a relevant language feature widely used in Java projects. Therefore, they can positively or negatively influence those systems' maintenance and evolution. Furthermore, these works also showed that code annotations might be misused in several ways, such as being misplaced \cite{Yu2019}, duplicated \cite{teixeira2018does, Yu2019}, and overused  \cite{Lima2018jss, Yu2019}. Consequently, developers might benefit from tools and approaches that aid in comprehending how code annotations are distributed in systems, and software visualization can be used for such purposes.
\section{Related Work}
\label{sec:related-work}

Researchers have explored software visualization to represent a target software system and improve code comprehension. These approaches usually use software metrics as the input data to create a graphical system representation. Both 2D and 3D approaches have been used, and even virtual reality-based visualizations. \textcolor{black}{However, designing a software visualization approach focusing on code annotations is a novel study to the best of our knowledge. For this reason, in this section, we focus on presenting the similarities between our design decisions and how other authors evaluated their approaches.}

Lanza and Ducasse~\cite{lanza2003} introduce the concept of a polymetric view, a lightweight software visualization enriched with software metrics. This visualization is based on shapes, such as rectangles, using metrics values to determine their size, color, and position. The polymetric view was also designed to represent relationships between entities. The represented entities in the software might have an edge connecting them to another entity, and the color and thickness of this edge might also represent some metric. This approach for visualization can support five metrics on each rectangle and two metrics on each edge. For the color, usually, the darker the color, the higher the metric value being represented. They developed a tool named CodeCrawler to create a polymetric view for a given system~\cite{lanza2004}, but it is currently unavailable. \textcolor{black}{As opposed to our visualization approach, we use only circles to represent the source code information. The reason is that we used the circle packing approach to aid in displaying the source code hierarchical structure.}  

Francese et al.~\cite{francese2016} propose a polymetric view for traditional and object-oriented metrics. Their work aims to provide an overview of the observed software in size, complexity, and structure. The view is based not only on static information but also on how classes exchange messages. This view is created as a \textcolor{black}{directed} graph where each basic unit is a ``type definition'' (class, interface, enum) that composes a graph as a node. Each node is drawn as an ellipsis and a rectangle, and their measures and colors represent several source code metrics.
This visualization was implemented in an Eclipse Plugin called MetricAttitude~\cite{francese2014}. To evaluate how their approach aided in Java software comprehension, they conducted a questionnaire-based experiment with 18 participants, ten undergraduate students and eight professional developers. The target software used was the JLine\footnote{\url{https://github.com/jline/jline3}}. Participants filled a \textit{comprehension questionnaire} with 16 open-ended questions about the target software system. Afterward, they filled out a \textit{perception questionnaire} about their impressions and opinions of the visualization tool. According to the author's report, the participants found the tool easy to understand and expressed an overall favorable judgment. As for the comprehension, the average correctness of answers has values between 0.70 and 0.87. 

Comparing the experiments carried out by Francese et al.~\cite{francese2016} with our approach to evaluating the CADV, we also conducted a questionnaire to find the usefulness and ease of use. However, since our approach focuses on visualizing code annotations, we did not find other software visualization tools or approaches to compare. As such, we compared our tool with manual code inspection aided by tools present in IDEs for searching in the project's source code. Nevertheless, we also conducted interviews with the developers and conducted qualitative data analysis (as described in Section~\ref{sec:avisu-experimenta-design}).

Another approach used to represent software systems is to create a synthetic natural environment (SNE). The goal is to create familiar environments artificially using metaphors, such as a city~\cite{Wettel2011}, a forest~\cite{erra2012}, and the solar system~\cite{graham2004}. \textcolor{black}{Then the system maps} software characteristics, mainly through metrics, to this environment. Possibly the most well-known SNE used for software visualization is the \textit{city metaphor} popularized by Wettel and Lanza~\cite{wettel2007city}, where the authors developed a tool called CodeCity~\cite{Wettel2011} to implement this metaphor. The \textit{city metaphor} represents types (classes, interfaces, enums) as buildings (parallelepipeds). In turn, these buildings are localized inside districts, which represent packages. The visual properties of each parallelepiped represent the software metrics of the class. \textcolor{black}{One aspect in common with these SNE is that they were all designed to use traditional source code metrics to display the system, in contrast with ours that use code annotation metrics.}

Wettel et al.~\cite{Wettel2011} performed an experiment to assess the CodeCity and \textit{city metaphor}. They elaborated nine questions that the participants should answer using CodeCity. For instance, one question was to \textit{Locate all the unit test classes}, and another \textit{Find three classes with the highest number of methods (NOM)}. The authors obtained an increase of (+24\%) task correctness and a decrease in the completion time (-12\%) compared to traditional tools such as IDEs. CodeCity constantly outperformed manual code inspection in tasks that benefit from an overview of the system. Comparing these results with our evaluation (discussed in Section~\ref{sec:discussion}), we also observed good performance of questions related to a general view of the system. 

The city metaphor gained a Virtual Reality (VR) version~\cite{merino2017}. The authors proposed the CityVR - an interactive visualization tool that uses virtual reality to implement the city metaphor. The authors conducted a qualitative empirical evaluation using semi-structured interviews with six experienced developers. From their results, developers felt curious and spent considerable interaction time navigating and selecting elements. Finally, the participants were willing to spend more time using CityVR to solve software comprehension tasks. \textcolor{black}{Compared with our work, we conducted a similar evaluation by interviewing six developers. Nevertheless, we also conducted a study with students to obtain a complementary result.} 

More recently, a new tool called Code2City was developed to support VR and flatscreen visualization~\cite{romano2019}. The authors conducted experiments with 42 participants comparing three different approaches: Code2City displayed on (i) a regular computer screen, (ii) the VR version, and (iii) a plugin for the Eclipse IDE named \textit{Metrics and Smells} that collects metrics and detects bad smells. The authors concluded that the city metaphor increases software comprehension, and users using the VR version concluded the experiment tasks more quickly and were more satisfied. 

Finally, Merino et al.~\cite{merino2018} mention that several software visualizations work has failed in their evaluations. Therefore, they provide little evidence of the effectiveness of their approach. They mention that 62\% of the proposed approaches do not include any evaluation or a weak one, such as anecdotal evidence of simple usage scenarios. They conclude that software visualization research should use more surveys with the target audience (to extract helpful information), conduct interviews, and perform an experiment using real-world software systems and a controlled system with practical tasks. These can be viable options to evaluate and assess a software visualization approach. As mentioned, we conducted an empirical evaluation with students and developers to support our findings to overcome this.

In conclusion, the visualizations we found in the literature focus on displaying size, complexity, and cohesion, with the types (classes, interfaces, enums) representing the main elements. Even though annotations are a relevant code element in Java, we did not find any software visualization approaches focusing on them. Since most visualization focuses on class structure metrics, they do not display much information about class responsibilities. For instance, in the study performed with CodeCity~\cite{Wettel2011}, to complete the task of locating unit test classes, the user should search for classes (or packages) containing the word ``test''. On the other hand, code annotations are usually tied to a specific responsibility or architectural role~\cite{aniche-emse-mvc-smells, aniche-satt-code-metric-thresholds}. Therefore, a visualization approach that allows identifying the annotation schemas present in classes and packages provides valuable information to identify how the responsibilities are organized in the system under analysis.
\section{Research Design}
\label{sec:rd-visu}

This section presents the research design to propose and evaluate the CADV approach. To build our visualization, we follow best practices and guidelines applied in previous works about software visualization~\cite{francese2016,merino2018,romano2019}. The following subsections present and discuss the goals for CADV, the steps to reach them, the target audience, and our approach for empirical evaluation based on questionnaires and interviews. In short, we took the following steps to reach our goals:

\begin{itemize}
    \item Proposed and designed the visualization approach for code annotation distribution named CADV;
    \item Developed a tool, named AVisualizer, that generates CADV for a given system and provides navigability through the different views;
    \item Applied questionnaires and conducted interviews to evaluate CADV through the use of AVisualizer, aiming to assess its effectiveness for program comprehensions tasks, usefulness, and ease of use;
    \item Analyzed the data collected from the questionnaires and interviews, combining quantitative and qualitative strategies to evaluate the proposed approach, raising its applicability, strengths, and weaknesses.
\end{itemize}

\subsection{Visualization Goals}
\label{subsec:visu-goals}

The primary goal of CADV is to aid in the comprehension and provide an overview of how code annotations are distributed in the analyzed software system. Then, the visualization should also provide more details of code annotations in specific classes and packages that the user wishes to explore further. Following, we describe the five goals.

\begin{itemize}

    \item [(\#G1)] - \textbf{Detect annotations schemas and how they are distributed in the packages}:  The user should be able to spot all schemas and quickly identify where they are being used. For instance, if the code annotations from a given schema are present in different packages, it might indicate that the responsibilities related to that schema are distributed in these packages. As another example, a concentration of an annotation schema usage in a package enables the user to locate the classes that handle the respective responsibility quickly. To reach this goal, the view should represent the annotations from each schema in the whole system and should not overwhelm the user with details closer to the source code.

    \item [(\#G2)] - \textbf{Detect how annotations are distributed per class in packages}: From the general view of the system, the user might want to investigate specific packages to obtain more details about code annotations in its classes. For instance, (i) how schemas are distributed inside classes, (ii) if the classes are coupled to several schemas, and (iii) if there are classes with a large number of annotations. An annotation size metric, such as LOCAD (Lines of Code in Annotation Declaration), can represent the annotations. To reach this goal, the view should present the code annotations characterized by a size metric grouped by their classes.

    \item [(\#G3)] - \textbf{Detect how annotations are distributed and grouped per code elements inside the classes}: Code annotations are placed on code elements, such as methods, members, and type definitions. Several annotations can be added to the same code element. The user might be interested in how the annotations are grouped inside these code elements to identify (i) if a specific code element is overloaded with annotations, (ii) if several code elements contain repeated annotations, and (iii) how annotations from different schemas are distributed inside the class. The size of the annotations should also be represented by a metric that can characterize its size. The view should group the annotations by code elements inside the classes to reach this goal.

    \item [(\#G4)] - \textbf{Provide a navigation system between views with different granularity}:  For each goal presented previously,  \#G1,  \#G2, and  \#G3, we are proposing a different view consistent with each of these three goals. Therefore, our visualization approach should provide some mechanism to navigate these different views.
    
    \item [(\#G5)] - \textbf{Detect misconfigurations}: Our goals are concerned with the distribution of code annotations in the system rather than detecting problems related to them. However, we might also detect potential problems if we can visualize code annotations. \textcolor{black}{According to Yu et al.~\cite{Yu2019}, code annotations are deleted because they are redundant (15.4\%) or wrong (24.5\%). The same study also points out that 19.8\% of the annotation replacements switch them to the ``same name" annotations from other libraries, revealing that annotation from an unexpected schema was present. The proposed approach should identify inconsistent annotations, enabling the user to locate potentially misplaced annotations.}

\end{itemize}

\subsection{Target Audience}

Software visualization approaches should clearly state the target audience, which should be consistent with the goals. Following, we describe the target audience aimed for the visualization approach: 

\begin{itemize}

    \item Newcomer Developer\footnote{The term "newcomer developer" is used in this work to refer to someone who has just joined a team. It does not necessarily mean that he/she has no software experience since even a senior software developer can be a newcomer.}: According to Dagenais et al.~\cite{Dagenais2010}, whenever a developer approaches a new software system, he/she feels like an explorer who need to orient themselves within an unfamiliar landscape, even senior developers. The goal \#G1 provides a general overview of the system that may help newcomers.
    
    \item Student: Given that they are constantly learning, they are also part of the target audience. They could use the visualization like newcomer developers to understand better some concepts related to code annotations and metadata-based frameworks usage. They can also identify aspects of the software system and further study them using other methods.
    
    \item Software Architect: Since he/she already has a firm grasp of the whole system, this audience can use the visualization approach for a more detailed analysis focusing on aspects such as: how to make this system better adhere to the proposed software architecture? Are annotations growing out of control in some parts of the system? Are package responsibilities clearly defined? Are there any misplaced or misconfigured code annotations? Are there any packages coupled with different annotation schemas?  
    
\end{itemize}

The target audience does not need to adhere strictly to these definitions. A newcomer can use the visualization to detect outliers, just as a seasoned developer can also gain new insight about the system software it did not know.

\subsection{Evaluation and Data Analysis Approach}
\label{sec:avisu-experimenta-design}
\label{sec:targetsfw}

\textcolor{black}{The CADV, as further explained in Section \ref{sec:cadv}, uses values extracted from the suite of metrics dedicated to code annotations~\cite{Lima2018jss}. Given the novelty of these metrics, we were unable to find a tool to compare with our solution. Therefore, to evaluate the CADV, we conducted an empirical study using questionnaires and interviews where participants had to use the AVisualizer, inspect a target software system, and answer some questions. We assessed the effectiveness of the tool in (i) program comprehension tasks related to code annotations, (ii) its perceived usefulness, (iii) its perceived ease of use, and (iv) the audience that it can potentially benefit.}

\textcolor{black}{The program comprehension tasks were elaborated based on real-world scenarios extracted from our previous experience~\cite{guerra2010,guerra2016annotationapis,teixeira2018does,Lima2018jss,Lima2020}, our visualization goals (Section~\ref{subsec:visu-goals}), and also tasks other researchers elaborated when evaluating their software visualization approach~ \cite{Wettel2011,francese2016,romano2019}}

\textcolor{black}{The correctness of the program comprehension tasks was defined by three authors, all with previous experience with code annotations. They were not involved in the development of the target system nor had any direct relationship with the developers. To elaborate on the tasks, they were granted access to the code. After the interview was carried out, all six developers agreed with the correctness of the answers.} 

The categories \textbf{perceived ease of use} and \textbf{perceived usefulness} were based on the TAM (Technology Acceptance Model)~\cite{davis1989}. These two variables characterize how users find the technology easy to use and how users find the technology helpful. The questions we elaborated to measure them used the Likert scale, ranging from \textit{strongly disagree} to \textit{strongly agree}, and a strategy similar to the work of Choma et al.~\cite{choma2019}.

We generated the proposed view for a real-world system to perform the evaluation, using a module from EMBRACE as the target software. It is a web application whose goal is to show space weather information developed by the National Institute for Space Research (INPE)\footnote{\url{https://www.gov.br/inpe/} - partially available in English}, a Brazilian public research institution active in the fields of meteorology and aerospace. INPE has the mission to foster scientific research, technological applications, and qualifying personnel in space and atmospheric sciences, space engineering, and space technology. It is vital in environmental protection, monitoring forest fires, and deforestation. The institute requires enterprise software systems that allow processing, persisting, and making data available to the community. This work was developed in INPE, and using a system in production from this institute as a target from our study applies our approach in a real-world application developed to meet the demands of scientific research institutions.

The \textbf{EMBRACE} web application is composed of several modules that follow the same reference architecture~\cite{SANTANA2014} based on Java Enterprise APIs. The application is used to process and make data publicly available. This software system has 1314 classes, from which 837 (64\%) contain at least one annotation. The application comprises six modules spanning 94 components archived in independent deployment units. We selected the module \textbf{SpaceWeatherTSI} to be used as the target software in the evaluation, given that it uses annotations from different metadata-based frameworks. This attribute matches the characteristics of a software system that could benefit \textcolor{black}{from the CADV approach,} that is currently in production. \textcolor{black}{Afterward, we conducted an interview study with six developers involved in constructing the target software and a questionnaire study with 79 master and undergraduate students from three different institutions. The idea of applying a questionnaire and interviews with different groups have been used in other Software Engineering research~\cite{wen2020, murphyHill2014}. We aim to capture different perspectives from groups with different backgrounds and experiences. The design of this study is described in more detail in Section~\ref{sec:evaluation}.}

Afterward, we conducted qualitative and quantitative analyses of the data obtained from the abovementioned studies. With this information, we evaluated the approach's effectiveness based on the number of correct answers to the program comprehension questions and how useful, and easy the tool was perceived. Finally, since the interview was semi-structured and the questionnaire also provided one open-ended question, we conducted a qualitative analysis to extract relevant points and insights about the proposed approach. 
\section{Code Annotations Distribution Visualization - CADV}
\label{sec:cadv}

This section presents the CADV proposal and definition. It comprises three different circle packing views, where a leaf node size is calculated based on an annotation-metric value. The tool that implements the CADV supports a navigation system that allows the user to switch between these three views.

The CADV displays the annotations used in a given class or package context. Given the hierarchical structure inherent to a source code, we chose a circle packing approach as a basis for the visualization. According to Bostock~\cite{d3CirclePack}, even though a circle packing is not space-efficient, it can better reveal the hierarchical structure.

Source code information such as packages, classes, and annotations are displayed as circles. We also use colors and outlines to differentiate them. Some elements are only visible in specific views, and the hierarchical structure of the source code also organizes the elements since code annotations are placed inside a class and classes inside packages.

To guide the design process of the views, we used a GQM (Goal Question Metric) Model~\cite{Caldiera1994}. From the five goals (Section \ref{sec:rd-visu}), four questions were extracted to be used in the model. Table \ref{tab:gqm-visualization} presents the GQM model with the questions and the respective goal from which they were extracted. We deliberately left \#G4 out of the GQM model since it is related to the navigation system, which had become a requirement for the tool implemented later.

\begin{table}[h]
\normalsize
\centering
\caption{GQM applied for the CADV Approach.}
\label{tab:gqm-visualization}
\begin{adjustbox}{max width=\linewidth}
\begin{tabular}{lll}
\hline
\textbf{Goal}   & (Purpose)		& \textit{Visualize} \\
         	& (Issue) 		& \textit{the usage and distribution of} \\
         	& (Object)	 	& \textit{annotated code} \\
         	& (Viewpoint) 		& \textit{from software developer viewpoint} \\
         	&  			& \\
\hline
(Question) 	& Q1			& \textbf{How are annotations schemas distributed by packages? (Extracted from G1)} \\
\hline
         	&  			& \\
(View 1)      & \textbf{System View}		& Provides a polymetric view that displays annotation schemas being used by packages \\\\
\hline
(Question) 	& Q2 			& \textbf{How are annotations schemas distributed inside packages? (Extracted from G2)} \\
\hline
         	&  			& \\
(View 2)     	& \textbf{Package View} 		& Provides a polymetric view that displays annotations being used in classes of a package\\\\

\hline
(Question) 	& Q3  			& \textbf{How are annotations schemas distributed inside classes? (Extracted from G3)} \\
\hline
         	&  			& \\
(View 3)    	& \textbf{Class View}		& Provides a polymetric view that displays annotations, and how they are grouped in the class code elements.\\\\
\hline
(Question) 	& Q4 			& \textbf{How to detect potential misplaced code annotations? (Extracted from G5)} \\
\hline
         	&  			& \\
(View 1)     	& \textbf{System View}		& -- \\
(View 2)     	& \textbf{Package View}		& -- \\
(View 3)     	& \textbf{Class View}		& -- \\\\
\hline

\end{tabular}
\end{adjustbox}
\end{table}

Every view may contribute to reaching the goal \#G5 differently and within their scope. For instance, the \textbf{Package View} seems suited to detect an extensive annotation. In contrast, the \textbf{System View} does not help much since it does not display any information about the annotation size. On the other hand, an annotation schema potentially misplaced might be quickly spotted in the \textbf{System View}. The \textbf{Class View} may contribute to detecting a specific code element overloaded with code annotations or annotations in unexpected code elements.

The CADV is a software visualization approach that different tools can implement. We developed AVisualizer as a reference implementation for the proposed approach, a web application that generates CADV, allowing users to interact with the views and navigate between them. In the following sections, we present CADV through the usage of AVisualizer. That way is simple to explain the static structure of the views and their interaction with the user. This tool was used in the studies performed to evaluate the proposed approach. The D3.js library \cite{Bostock2011} was used as the basis to develop the AVisualizer.

The following subsection describes the basic layout of the AVisualizer tool, and the subsequent subsections define the three views that compose the CADV. Reference images of the views were generated by the AVisualizer tool using the \textit{SpaceWeatherTSI} as the target software (described in Section~\ref{sec:targetsfw}). Following the same guidelines as Lanza and Ducasse~\cite{lanza2003} and Romano et al.~\cite{romano2019}, we recommend reading this work on a colored screen monitor. In our supplementary material~\cite{zenodo} there are open-source projects used as an example for the AVisualizer. The demonstration of the AVisualizer is available at \url{https://avisualizer.herokuapp.com/}

\subsection{AVisualizer Layout}

\begin{figure}[h]
    \centering
    \includegraphics[width=\linewidth]{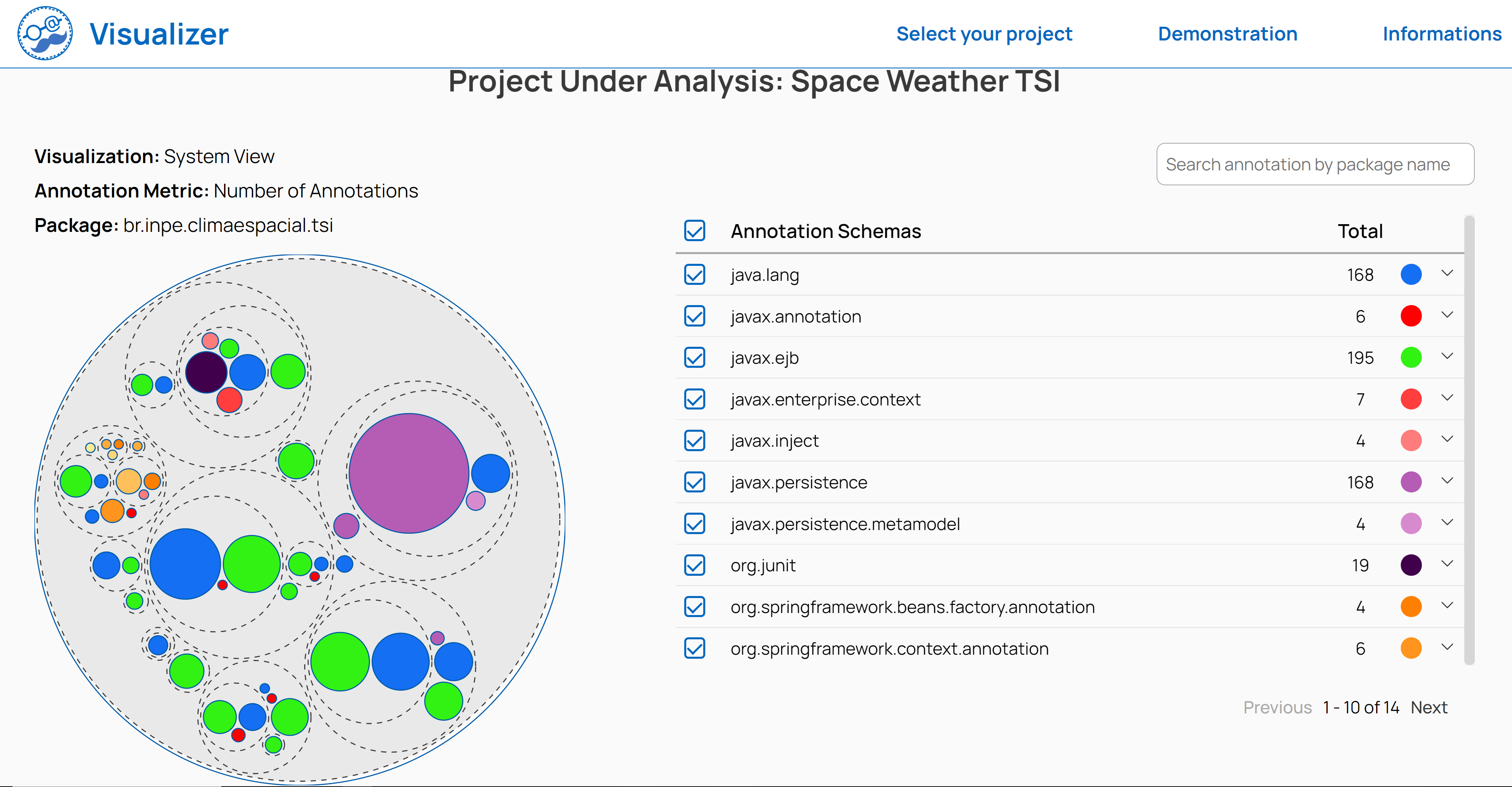}
    \caption{AVisualizer tool with System View}
    \label{fig:avisu-cadv}
\end{figure}

Figure~\ref{fig:avisu-cadv} presents the initial page of the  AVisualizer tool. The circle packing view, drawing a given system under analysis, is presented on the left side. This interface is constantly updated as the user navigates the tool. Above the view, there is a small header with the following information:

\begin{itemize}
    \item Visualization: Informs which of the three views is currently rendered. It could be \textbf{System View}, \textbf{Package View}, or \textbf{Class View}. 
    \item Annotation Metric: Inform the metric value used to render the leaf nodes.
    \item Package or Class: Informs what package or class is currently being displayed, represented by the outermost circle. 
\end{itemize}

On the right of the visualization, the table \textbf{Annotation Schemas} lists all schemas found on the project, displaying the associated color and informing their total number of annotations. For instance, there are 168 annotations from the \textbf{javax.persistence} schema, and 19 from the \textbf{org.junit} schema. By associating a color to a schema, the user can spot its annotations by searching circles of that color inside circles representing packages or classes. There is also a check box on the table that allows hiding specific annotations.

AVisualizer uses fixed colors for some common annotation schemas found in open-source software~\cite{Lima2018jss}. This strategy helps to standardize the usage of the AVisualizer since common schemas will receive the same color in the views generated for different target systems. For instance, given that pink is used for the \textbf{javax.persistence}, regardless of the software being analyzed, if the user spots a pink circle, she will associate it with persistence. If the schema found in the system is not predetermined, then the AVisualizer randomly associates one. However, white and gray colors do not represent schemas since they already have other meanings within the CADV. The following is a list with some examples of these predetermined schemas:

\begin{itemize}
    \item \texttt{java.lang} - Blue  
    \item \texttt{javax.persistence} and \texttt{org.hibernate} - Pink tones
    \item \texttt{org.springframework} - Orange tones
    \item \texttt{org.junit} and \texttt{org.mockito} - Purple
    \item \texttt{javax.ejb} - Green
\end{itemize}

Related schemas will use a tone variation of the same color. For instance \textbf{javax.persistence} and \textbf{javax.persistence.metamodel} will have different pink tonality. Finally, the views do not display any textual information to create a clean visualization. However, the user can hover the mouse over the circles to reveal labels with more information. 

\subsection{System View}

The \textbf{{System View}}, which is presented in Figure \ref{fig:avisu-cadv} is the initial view displayed to the user. It displays the whole project allowing users to overview how code annotations are spread in the system under analysis. The \textbf{System View} displays circles representing packages and a representation of the annotation schemas usage. The light gray background color was chosen to remain as neutral as possible. The following list presents the characteristics of these circles and what information becomes available when the user hovers the mouse over the circle representing the annotation.

\begin{itemize}[nosep]

    \item Packages: Every circle representing a package has a dashed outline. The outermost dashed circle represents the root package of the project. The inner ``dashed outlined circles'' are other packages, which can be at the same level or nested. In Figure \ref{fig:avisu-cadv}, there are several packages in this system, each represented by circles with a dashed outline. This approach displays the hierarchical structure of the source code under analysis. The user can also click on these circles (packages) to perform a zoom action. When this happens, the outermost circle changes, and the header will reflect the current package. 
    
    \item Annotation Schemas Usage: These are colored filled circles rendered inside ``dashed outline circles'' (packages). They represent the occurrence of annotations of the schema mapped to the circle color inside that package. The \textbf{Annotation Schemas} table works as a legend for the annotation schema color. The size of these circles is proportional to the number of code annotations of that particular schema inside that package. The larger the circles, the higher the number of annotations from that schema. For instance, the \textbf{System View} on Figure \ref{fig:avisu-cadv} shows a large pink-toned circle, meaning that this package has a high number of annotations from the \textit{javax.persistence} schema. 
    
    \item \textcolor{black}{Label: When the user hovers the mouse over a colored circle, i.e., an annotation, a label appears with the following information: (i) the name of the schema, (ii) the name of the package and, (iii) the number of occurrences of annotations from this schema in the package.}  
\end{itemize}

In the \textbf{{System View}}, the classes of the packages are not visible. It does not show in how many classes the annotations are divided. To access this information, the user should click on the package to navigate to the \textbf{{Package View}}, described in the next section.

\subsection{Package View}

The \textbf{Package View} can display classes and individual code annotations inside a given observed package. Differently from the \textbf{System View} designed to visualize the whole system, the \textbf{Package View} displays a specific package. Figure~\ref{fig:avisu-pacakge-view} displays an example of the \textbf{Package View}. The circles are rendered with the following characteristics: 

\begin{figure}[ht]
    \centering
    \begin{subfigure}[t]{0.5\textwidth}
        \centering
        \includegraphics[width=0.95\linewidth]{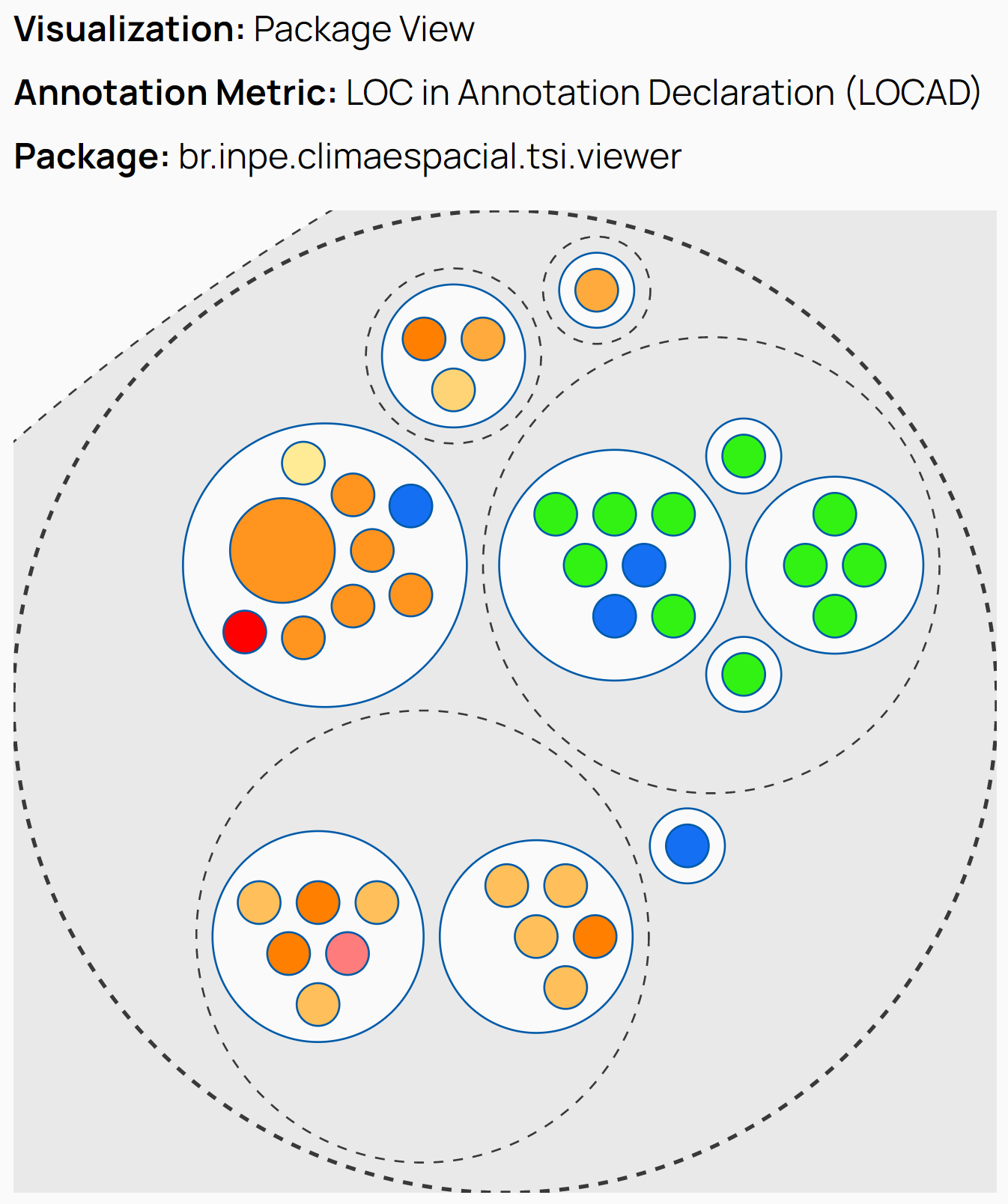}
        \caption{Package View}
        \label{fig:avisu-pacakge-view} 
    \end{subfigure}%
    ~ 
    \begin{subfigure}[t]{0.5\textwidth}
        \centering
        \includegraphics[width=0.95\linewidth]{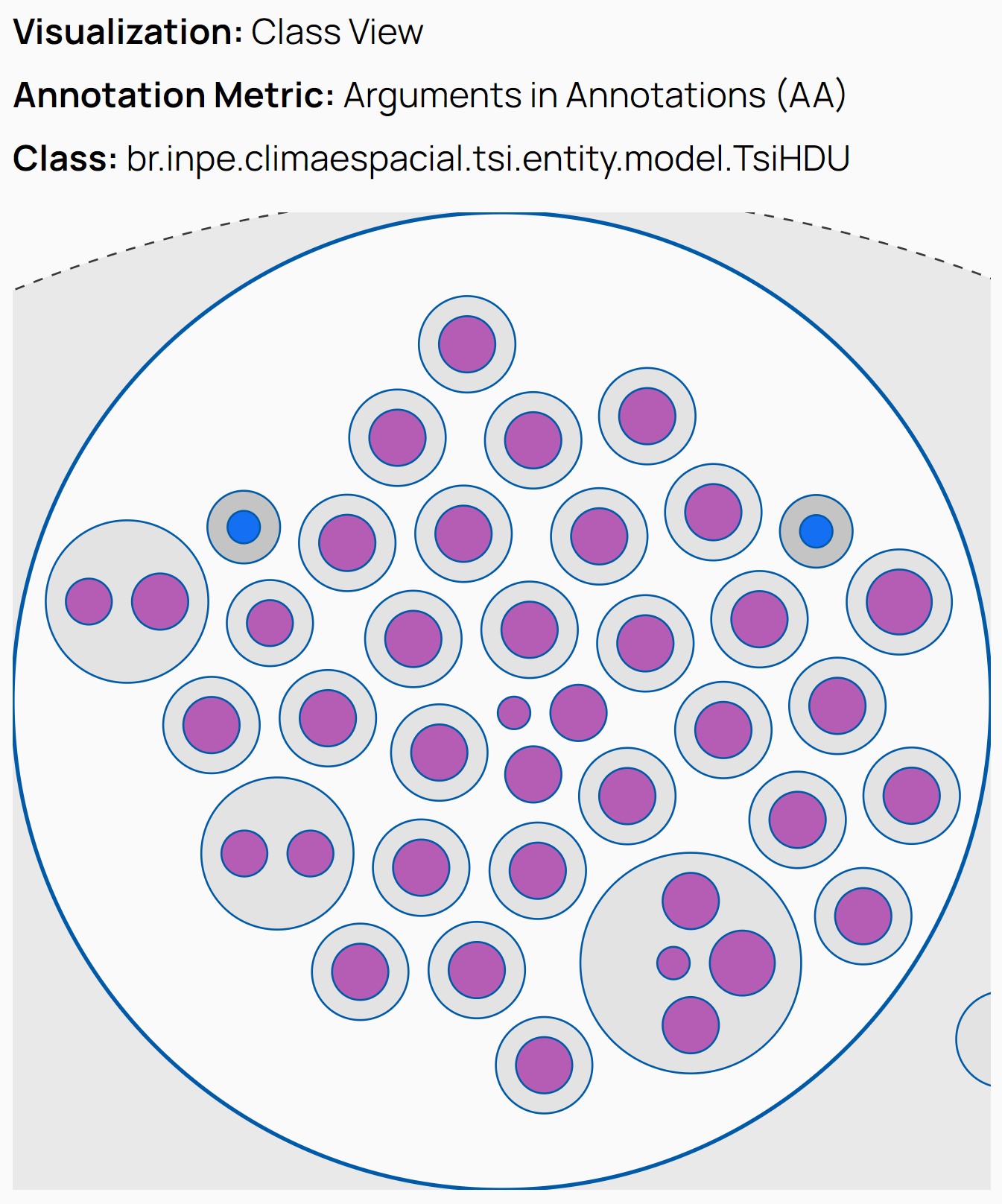}
        \caption{Class View}
        \label{fig:avisu-class-view} 
    \end{subfigure}
    \caption{An example of the Package and Class View}
\end{figure}

\begin{itemize}
    \item Package: It has the same characteristics as the \textbf{System View}.
    \item Classes: Classes are rendered as white-filled circles. Their size depends on the number of code annotations used inside the class. If a white circle appears large than other white circles, it represents a class with more code annotations.
    \item Code Annotations: These are colored (any color besides white and gray) filled circles rendered on top of white circles. They represent code annotations being used inside a specific class. Their color matches the color of their schema, present on the \textbf{Annotations Schema} table. The size of these circles is proportional to their LOCAD value, i.e., the default metric used in the \textbf{Package View}.
    \item \textcolor{black}{Label: When the user hovers the mouse over a colored circle, i.e., an annotation, a label appears with the following information: (i) the name of the package, (ii) the name of the class, (iii) the name of the annotation and, (iv) the metric used to determine the size of the circle. For the Package View, the default metric is the LOCAD (LOC in Annotation Declaration).}  
\end{itemize}

The \textbf{{Package View}} does not display how individual annotations are grouped in a specific code element inside a class. To access this information, the user should click on the class to navigate to the \textbf{{Class View}}. Comparing the \textbf{Package View} with the \textbf{{System View}}, it displays information with finer granularity but for a more restricted context.

\subsection{Class View}

The \textbf{Class View} displays classes with individual code annotations inside the observed class. It shows how code annotations are distributed by code elements, such as a method, a field, or the class itself. Figure \ref{fig:avisu-class-view} presents an example of \textbf{Class View}. The circles are rendered according to the following rules:


\begin{itemize}[nosep]
    \item Classes: Just as in the \textbf{{Package View}}, they are rendered as a white circle. There is only one white circle enclosing all the others since the focus is on a specific class.
    \item Annotations: Annotations are represented individually by colored circles, whose color represents the schema. The circle size is based on the AA (Arguments in Annotations) metric. 
    \item Gray-Circles: This color is also used to represent packages, but in the \textbf{{Class View}} they represent code elements, such as method and class members. The code annotations (colored circles) are rendered on top of these gray circles. Colored circles rendered directly on top of a white circle represent code annotations configuring the class itself. The number of colored circles rendered on the same gray circle represents the AED (Annotations in Element Declaration) metric.
    \item \textcolor{black}{Label: When the user hovers the mouse over a colored circle, i.e., an annotation, a label appears with the following information: (i) the name of the package, (ii) the name of the class, (iii) the name and type of the element the annotation is configuring, (iv) the name of the annotation and, (v) the metric value used to determine the size of the circle. For the Class View, the default metric is the AA (Arguments in Annotation).} 
\end{itemize}

The gray-toned background used to represent code elements is the same one used to represent a package background in the other two views. However, it does not affect the \textbf{{Class View}} since it shows a single class, and no package is displayed. Furthermore, the gray color provides a suitable neutral background. Apart from white and gray, every other color represents a schema.

Inspecting the \textbf{Class View}, it is unclear whether the gray circle represents a method, enum, or other code elements. This design decision was made to avoid overloading users with more information. Users, however, can access this information by pointing the mouse to it. The \textbf{Class View} can display the grouping of code annotations by code elements based on the size of these gray circles. Furthermore, the schema information is always available since each one has a unique color throughout the project. 

The metric used to render the leaf nodes, i.e., the colored circles representing individual annotations, was the AA (Arguments in Annotations). In other words, annotations with more arguments are larger. We decided to use this metric to vary from the one used in another view since \textbf{Package View} already uses LOCAD.

\subsection{Code Annotations Distribution Visualization Summary}

The CADV is a circle packing-based view designed to aid in visualizing code annotations used in software systems. It comprises three different views to display information at different granularity levels. This approach allows users to switch between these views and analyze the parts of the software with the desired granularity:

\begin{description}

\item[System View (SV):] Related to the goals G1 and G5, displays packages and schemas used. Each schema is assigned a color, and packages are gray circles with a dashed outline.

\item[Package View (PV):] Related to the goals G2 and G5, displays a package, classes, and code annotations used in the classes. Classes are white circles. Code annotations are assigned the schema color and are rendered on top of the white circles. Packages are gray circles with a dashed outline.

\item[Class View (CV):] Related to the G3 and G5, displays a class, code elements, and code annotations grouped by code elements. Classes are white circles. Code elements are a gray color. Code annotations are assigned the schema color and are rendered on top of either the white circles or gray circles.

\end{description}

\section{Empirical Evaluation Design}
\label{sec:evaluation}

This section describes the evaluation approach conducted for CADV using the AVisualizer tool. The following subsections present the steps of the interview, the questionnaires, and the qualitative analysis.

\subsection{Interview}

We invited developers who worked in the \textbf{SpaceWeatherTSI} (described in Section \ref{sec:targetsfw}) web application to participate in the interview study. Members involved in this project's construction could improve the assessment of the visualization approach. Given that they know the system, they could assess if the AVisualizer tool could effectively display a coherent view of the project under analysis. They could also analyze if the CADV brought new insights and valuable information to an already familiar system.

\subsubsection{Procedure}

We first provided a link with a recorded video as a tutorial on using the AVisualizer tool. We sent it to each participant 1 week before the interview. We also sent a Google Form link to collect personal data, programming background, and consent to participate in the evaluation. 

The interview occurred according to the following steps: 

\begin{enumerate}
    \item Video call: Using Google Meet.
    \item AVisualizer Demo:  We provided the link to the deployed demo of the AVisualizer tool to the interviewee.
    \item Recording: Initiated the recording using the OBS desktop application.
    \item Training Session: The interviewer shared the screen with the interviewee and provided another 15-minute training. Therefore, participants were submitted to two training sessions on the AVisualizer tool. 
    \item AVisualizer Execution: The interviewees could use the tool themselves (on their computer), but they preferred to have the interviewer manipulate it for them (in the interviewer's computer, with the screen shared). The interviewees issued all the commands while answering and discussing the questions. \textcolor{black}{In parts of the interview, the interviewees also interacted with the tool on their computer while providing feedback. This approach was natural since the demonstration tool was available as a web application and was easily accessed from a browser.}
    \item Questions: The interviewer started asking questions presented in Table~\ref{tab:comprehension-questionaire} to guide the interview. However, the conversation was informal, so the interviewees could freely answer and discuss other topics as the interview carried on.
\end{enumerate}

On average, the interviews lasted 60 minutes, i.e., 15 minutes of training with 45 minutes of questions/discussions.

\subsubsection{Interview Questions}

To guide the interview, we elaborated 15 questions presented in Table \ref{tab:comprehension-questionaire}. They were based on the four categories considered in the evaluation of the CADV using AVisualizer (discussed in Section~\ref{sec:rd-visu}). Since this was a semi-structured interview, we opted to elaborate questions that could tackle more than one category at once, making the interview more fluid. For instance, the first five questions directly ask the participants to execute program comprehension tasks in which the perceived ease of use can also be evaluated. Similarly, the following seven questions use the same strategy for the perceived usefulness.

\begin{table}[!ht]
\huge
\caption{Interview Questions.}
\begin{adjustbox}{width=\linewidth}
\begin{tabular}{lll}
\hline
\textbf{{{ID}}} & \textbf{Question}                                                                                                                                                                                           & \textbf{Goal} \\ \hline
\multicolumn{3}{c}{\textbf{Program Comprehension and Perceived Ease of Use}}                                                                                                                                                                                                                                            \\ \hline
Q1                                             & What annotation schemas are concentrated in fewer packages?                                                                                                                                                                  & G1                             \\
Q2                                             & What annotations schemas are present in more packages?                                                                                                                                                                       & G1                             \\
Q3                                             & \begin{tabular}[c]{@{}l@{}}Is the circle packing able to represent the hierarchical \\ structure of packages adequately?\end{tabular}                                                                                        & G1                             \\
Q4                                             & \begin{tabular}[c]{@{}l@{}}When changing to the \textbf{\textit{Package View}} (any package), \\ can you tell which class(es) contain the largest number of code annotations?\end{tabular} & G2/G4                          \\
Q5                                             & \begin{tabular}[c]{@{}l@{}}When changing to the \textbf{\textit{Class View}} (any class), \\ can you tell which code elements contain the largest number of annotations ?\end{tabular}     & G3/G4                          \\ \hline
\multicolumn{3}{c}{\textbf{Program Comprehension and Perceived Usefulness}}                                                                                                                                                                                                                                             \\ \hline
Q6                                             & Which package(s) contain model classes mapped to databases?                                                                                                                                                                  & G2                             \\
Q7                                             & Which package(s) contain web controllers classes?                                                                                                                                                                            & G2                             \\
Q8                                             & Which package(s) contain unit testing classes? Is there enough unit testing code?                                                                                                                                            & G2                             \\
Q9                                             & Is it possible to identify packages with specific responsibilities by visualizing schemas?                                                                                                                                   & G2                             \\
Q10                                            & Is it possible to detect potentially misplaced code annotations? Describe the steps                                                                                                                                          & G5                             \\
Q11                                            & Is it possible to detect code annotations potentially being used excessively?                                                                                                                                                & G5                             \\
Q12                                            & Out of the three views, System, Package, and Class, which one did you prefer?                                                                                                                                                & G1-G5                          \\ \hline
\multicolumn{3}{c}{\textbf{Target Audience}}                                                                                                                                                                                                                                                                            \\ \hline
Q13                                            & \begin{tabular}[c]{@{}l@{}}Do you believe the tool eased the process of seeing how the \\ code annotations are distributed in the system?\end{tabular}                                                                       & G1-G5                          \\
Q14                                            & Do you believe a newcomer developer could benefit from such a tool?                                                                                                                                                          & G1-G5                          \\
Q15                                            & What role in a software team do you think can better use the AVisualizer tool?                                                                                                                                               & G1-G5                         
\end{tabular}
\end{adjustbox}
\label{tab:comprehension-questionaire}
\end{table}

As explained in Section \ref{sec:dataanalysis}, we performed a qualitative analysis of the interview transcriptions to extract themes related to the target categories, such as ``perceived ease of use'' and ``perceived usefulness''. For the program comprehension questions, the participants' answers were analyzed and classified as ``right'' or ``wrong''.

\subsection{Questionnaire}

We invited students from Inatel\footnote{\url{https://inatel.br/home/}} (National Institute of Telecommunications) and IPT\footnote{\url{https://www.ipt.br/}} (Institute for Technological Research) in Brazil, and UniBz\footnote{\url{https://www.unibz.it/}} (Free University of Bozen-Bolzano) in Italy to answer the questionnaire. Seventy-nine students participated in the study, where 70\% were undergraduates in Computer Science/Engineering courses, and the remaining were master students. These studies were carried out remotely for eight months. None of these students was involved in developing the \textbf{SpaceWeatherTSI} software and provided a neutral point of view. Using students to conduct experiments and evaluations is a viable option to advance software engineering~\cite{falessi2018,host2000}. \textcolor{black}{The questionnaire was applied in the context of courses that explained concepts about code annotations for the students. Participation in the study was optional, no grade was associated with the answers, and we did not keep any information that could identify the students.}

\subsubsection{Procedure}

The questionnaire was carried out through a survey using Google Forms. As soon as the students accessed the form, they went through the following items:  

\begin{enumerate}
    \item Clarified Consent Term: A text explained the study's goals and requested their participation. The students could choose not to be involved.
    \item Personal Information: The questionnaire started with questions that gathered demographic information, experience with code annotations, current role, and primary programming language.
    \item Code Annotations: A text briefly explained code annotations and the code annotation metrics used in the views (Section~\ref{ssec:annotmetric}).
    \item AVisualizer Tutorial: The questionnaire provided a textual explanation of the AVisualizer tool and a link to a Youtube video tutorial of the tool. The participants were free to watch the video during the whole evaluation process. It was the same video used for the interview study.
    \item \textcolor{black}{Using the AVisualizer: The questionnaire provided a link to an instance of the AVisualizer tool displaying the views for the \textbf{SpaceWeatherTSI} system. While using the AVisualizer, the students had to answer 10  program comprehension tasks about this web application. 
    \item Impressions of the AVisualizer: After answering the 10 program comprehension tasks, they were presented with eight questions using the Likert Scale to measure the \textbf{perceived ease of use}. 
    \item Possible Scenarios Usage: To measure the \textbf{perceived usefulness} from the students, we presented a question with four possible scenarios and asked them which of these would most likely use the tool.
    \item Open-ended Question: Additionally, a final open-ended question asked the students to describe their overall opinion and impressions about the approach and the tool. The answer to this question was coded and used for the qualitative analysis.}
\end{enumerate}

\subsubsection{Program Comprehension}

We elaborated ten close-ended questions for the questionnaire with five alternatives each. These questions aimed to detect if the participants could identify code annotations characteristics from the target project using the AVisualizer tool. The questions were elaborated based on the goals proposed in Section \ref{subsec:visu-goals}. They can be found in our supplementary material~\cite{zenodo}.

\subsubsection{Perceived Ease of Use and Perceived Usefulness}
\label{sec:ease-use-usefulness-evaluation}

To measure the ``perceived ease of use'',  we elaborated eight statements that measured the students' impressions about the AVisualizer tool when navigating between the views and identifying packages and classes. These statements used the Likert Scale, ranging from \textit{strongly disagree} to \textit{strongly agree}. Table \ref{tab:tam-questionnaire} presents these statements.

\begin{table}[!ht]
\Large
\caption{Perceived Ease of Use - Statements}
\begin{adjustbox}{width=\linewidth}
\begin{tabular}{ll}
                                                     \\ \hline
\multicolumn{1}{l|}{\textbf{ID}} & \textbf{Statement}                                                                                            \\ \hline
\multicolumn{1}{l|}{SE1}                   & I can easily identify java packages with different responsibilities using the AVisualizer tool                \\
\multicolumn{1}{l|}{SE2}                   & I can easily see how code annotations are distributed in the system under analysis using the AVisualizer tool \\
\multicolumn{1}{l|}{SE3}                   & Learning how to use the AVisualizer was easy to me                                                            \\
\multicolumn{1}{l|}{SE4}                   & I can easily see how many annotation schemas are being used inside a java class using the AVisualizer         \\
\multicolumn{1}{l|}{SE5}                   & I can easily see how many annotation schemas are being used inside a java package using the AVisualizer       \\
\multicolumn{1}{l|}{SE6}                   & I can easily identify what java package I am currently inspecting using the AVisualizer                       \\
\multicolumn{1}{l|}{SE7}                   & I can easily identify the class I'm inspecting in the AVisualizer tool                                        \\
\multicolumn{1}{l|}{SE8}                   & I can easily navigate to and from the packages and classes being analyzed with the AVisualizer tool           \\ \hline
                                           &                                                                                                               \\\\

\end{tabular}
\end{adjustbox}
\label{tab:tam-questionnaire}
\end{table}

\textcolor{black}{To measure the ``perceived usefulness'' we presented the students with a question with four possible usage scenarios they would most likely use the tool: (i) Identify schemas to learn about them elsewhere; (ii) Search for large or misplaced annotations; (iii) analyze the architecture; (iv) Familiarize with the system before adding new features.}

\subsubsection{Target Audience}

The students were also asked what role in a software development team is the audience for the AVisualizer. They could check multiple options of the following roles: Developer, Architect, Tester, Quality Assurance, Project Manager, Framework Developers, and Newcomer Developer.

\subsection{Data Analysis}
\label{sec:dataanalysis}

We gathered the data obtained from the interviews and questionnaires to analyze and obtain the results considering the four categories mentioned in Section~\ref{sec:rd-visu}: (i) \textbf{Program Comprehension Tasks}, (ii) \textbf{Perceived Usefulness}, (iii) \textbf{Perceived Ease of Use}, and (iv) the \textbf{Target Audience} (iv).

The first item, \textit{Program Comprehension}, can be analyzed by validating the correctness of the answers. For the remaining three, we combined a qualitative and quantitative approach. The quantitative data can be extracted from the Likert Scale answers. The qualitative data collected in the interviews and answers to the open-question allowed us to obtain insights from the answers to the objective questions. To analyze this qualitative data, we followed a thematic analysis. This approach is ``a method for identifying, analyzing and reporting patterns (themes) within data''~\cite{Braun2006}. In this process, pieces of text are coded using labels that are later grouped in higher-order themes, increasing the abstraction level until a defined saturation point, e.g., a model, could be built. Therefore, we aim to label and categorize data using this approach to identify the strengths and weaknesses of the approach commonly mentioned by the subjects.

Cruzes and Dyb\aa~\cite{Cruzes2011} identify three ways a thematic analysis could be conducted. First, researchers build an initial list of codes that will guide them in the coding process in a deductive approach. Another option is an inductive approach where researchers do not have a starting list of codes and inspect the code systematically, e.g., sentence by sentence, to make some concepts emerge from data. Finally, the authors acknowledge an integrated approach that is a ``partway'' between the previous two. Using this approach, researchers have a general scheme pointing to some aspects, but the themes are built bottom-up. We followed an integrated approach in our analysis since TAM already gives us two themes: perceived ease of use and perceived usefulness. \textcolor{black}{Our supplementary material provides an example of the steps we used to extract some codes/themes as well as how they were consolidated/grouped in Table\ref{tab:coding}~\cite{zenodo}}.

\textcolor{black}{We hired an external service to transcribe the interviews. For the coding process, we employed NVivo 12\footnote{https://www.qsrinternational.com/nvivo-qualitative-data-analysis-software/home}. A researcher coded the transcriptions of the interviews and the answers to the open questions in the questionnaires. Then, another member of the authoring team reviewed the labels. The two researchers discussed the disagreements until they reached a consensus.} 
\section{Results and Discussion}
\label{sec:discussion}

This section presents our results and discussions from the conducted empirical evaluation with questionnaires and interviews. We follow the four categories to analyze the data as mentioned in Section \ref{sec:evaluation}. \textcolor{black}{Table \ref{tab:coding} presents the consolidated themes we obtained for \textbf{Perceived Ease of Use} and \textbf{Perceived Usefulness} by coding the interviews and the answers for the single open-ended question present in the questionnaire for students.} To differentiate the source, \textcolor{black}{we have two columns}, specifying the amount from the ``interviews'' and from ``questionnarie''. We used as labels: PES (``\textbf{P}erceived \textbf{E}ase of Use \textbf{S}trenghts''), PEW (``\textbf{P}erceived \textbf{E}ase of Use \textbf{W}eakness''), PUS (\textbf{P}erceived \textbf{U}sefulness \textbf{S}trengths) and PUW (\textbf{P}erceived \textbf{U}sefulness \textbf{W}eakness).

\begin{table}[!htb]
\caption{Themes}
\begin{adjustbox}{width=\linewidth}
\begin{tabular}{lllllccc}
\multicolumn{8}{c}{\textbf{\large Perceived Ease of Use - Themes}}                                                                                                                                                                                                                                                               \\ \hline
\multicolumn{1}{l|}{\textbf{Theme}} & \multicolumn{4}{c|}{\textbf{Strengths}}                                                                                                                & \multicolumn{1}{c|}{\textbf{Interview}} & \multicolumn{1}{c|}{\textbf{Questionnaire}} & \multicolumn{1}{l}{\textbf{Total}} \\ \hline
\multicolumn{1}{l|}{PES-1}          & \multicolumn{4}{l|}{Easy to see annotation's usage}                                                                                                    & \multicolumn{1}{c|}{0}                  & \multicolumn{1}{c|}{6}                      & 6                                  \\
\multicolumn{1}{l|}{PES-2}          & \multicolumn{4}{l|}{Easy to understand how to use the   tool}                                                                                          & \multicolumn{1}{c|}{2}                  & \multicolumn{1}{c|}{9}                      & 11                                 \\
\multicolumn{1}{l|}{PES-3}          & \multicolumn{4}{l|}{Easy to understand package hierarchy}                                                                                              & \multicolumn{1}{c|}{5}                  & \multicolumn{1}{c|}{0}                      & 5                                  \\
\multicolumn{1}{l|}{PES-4}          & \multicolumn{4}{l|}{Quick learning curve}                                                                                                              & \multicolumn{1}{c|}{0}                  & \multicolumn{1}{c|}{3}                      & 3                                  \\ \hline
\multicolumn{1}{l|}{\textbf{Theme}} & \multicolumn{4}{c|}{\textbf{Weakness}}                                                                                                                 & \multicolumn{1}{c|}{\textbf{Interview}} & \multicolumn{1}{c|}{\textbf{Questionnaire}} & \multicolumn{1}{l}{\textbf{Total}} \\ \hline
\multicolumn{1}{l|}{PEW-1}          & \multicolumn{4}{l|}{Understand the concept of annotation   schema}                                                                                     & \multicolumn{1}{c|}{8}                  & \multicolumn{1}{c|}{1}                      & 9                                  \\
\multicolumn{1}{l|}{PEW-2}          & \multicolumn{4}{l|}{Similar Colors may complicate the analysis}                                                                                        & \multicolumn{1}{c|}{3}                  & \multicolumn{1}{c|}{1}                      & 4                                  \\
\multicolumn{1}{l|}{PEW-3}          & \multicolumn{4}{l|}{Confusion when switching metrics   between views}                                                                                  & \multicolumn{1}{c|}{16}                 & \multicolumn{1}{c|}{1}                      & 17                                 \\
\multicolumn{1}{l|}{PEW-4}          & \multicolumn{4}{l|}{Understand the grouping in Class   View}                                                                                           & \multicolumn{1}{c|}{1}                  & \multicolumn{1}{c|}{2}                      & 3                                  \\
\multicolumn{1}{l|}{PEW-5}          & \multicolumn{4}{l|}{Confusing how to trigger a new view}                                                                                               & \multicolumn{1}{c|}{2}                  & \multicolumn{1}{c|}{3}                      & 5                                  \\
\multicolumn{1}{l|}{PEW-6}          & \multicolumn{4}{l|}{Headers sub-utilized}                                                                                                              & \multicolumn{1}{c|}{3}                  & \multicolumn{1}{c|}{0}                      & 3                                  \\
\multicolumn{1}{l|}{PEW-7}          & \multicolumn{4}{l|}{Tool requires training to use}                                                                                                     & \multicolumn{1}{c|}{3}                  & \multicolumn{1}{c|}{5}                      & 8                                  \\
\multicolumn{1}{l|}{PEW-8}          & \multicolumn{4}{l|}{Knowledge of Annotation Metrics}                                                                                                   & \multicolumn{1}{c|}{2}                  & \multicolumn{1}{c|}{1}                      & 3                                  \\ \hline
\multicolumn{8}{l}{}                                                                                                                                                                                                                                                                                                      \\
\multicolumn{8}{c}{\textbf{\large Perceived Usefulness - Themes}}                                                                                                                                                                                                                                                                \\ \hline
\textbf{Theme}                      & \multicolumn{4}{c|}{\textbf{Strengths}}                                                                                                                & \multicolumn{1}{c|}{\textbf{Interview}} & \multicolumn{1}{c|}{\textbf{Questionnaire}} & \multicolumn{1}{l}{\textbf{Total}} \\ \hline
\multicolumn{1}{l|}{PUS-1}          & \multicolumn{4}{l|}{\begin{tabular}[c]{@{}l@{}}Visualization is better than code inspection to \\ search annotations related information\end{tabular}} & \multicolumn{1}{c|}{10}                 & \multicolumn{1}{c|}{1}                      & 11                                 \\
\multicolumn{1}{l|}{PUS-2}          & \multicolumn{4}{l|}{More useful to medium and large projects}                                                                                          & \multicolumn{1}{c|}{0}                  & \multicolumn{1}{c|}{1}                      & 1                                  \\
\multicolumn{1}{l|}{PUS-3}          & \multicolumn{4}{l|}{Useful to improve and visualize the architecture}                                                                                  & \multicolumn{1}{c|}{4}                  & \multicolumn{1}{c|}{6}                      & 10                                 \\
\multicolumn{1}{l|}{PUS-4}          & \multicolumn{4}{l|}{Detection of frameworks that are interchangeable}                                                                                  & \multicolumn{1}{c|}{1}                  & \multicolumn{1}{c|}{0}                      & 1                                  \\
\multicolumn{1}{l|}{PUS-5}          & \multicolumn{4}{l|}{Detection of Responsibilities Using Schemas}                                                                                       & \multicolumn{1}{c|}{6}                  & \multicolumn{1}{c|}{1}                      & 7                                  \\
\multicolumn{1}{l|}{PUS-6}          & \multicolumn{4}{l|}{\begin{tabular}[c]{@{}l@{}}Color strategy helps identify potentially \\ misplaced or inconsistent annotation\end{tabular}}         & \multicolumn{1}{c|}{6}                  & \multicolumn{1}{c|}{1}                      & 7                                  \\
\multicolumn{1}{l|}{PUS-7}          & \multicolumn{4}{l|}{Update libraries and frameworks}                                                                                                   & \multicolumn{1}{c|}{2}                  & \multicolumn{1}{c|}{0}                      & 2                                  \\
\multicolumn{1}{c|}{PUS-8}          & \multicolumn{4}{l|}{\begin{tabular}[c]{@{}l@{}}Newcomer may familiarize with the system \\ before contributing\end{tabular}}                           & \multicolumn{1}{c|}{3}                  & \multicolumn{1}{c|}{1}                      & 4                                  \\ \hline
\multicolumn{1}{l|}{\textbf{Theme}} & \multicolumn{4}{c|}{\textbf{Weakness}}                                                                                                                 & \multicolumn{1}{c|}{\textbf{Interview}} & \multicolumn{1}{c|}{\textbf{Questionnaire}} & \textbf{Total}                     \\ \hline
\multicolumn{1}{l|}{PUW-1}          & \multicolumn{4}{l|}{I can find large annotations, but this is not very useful isolated}                                                                & \multicolumn{1}{c|}{12}                 & \multicolumn{1}{c|}{0}                      & 12                                 \\
\multicolumn{1}{l|}{PUW-2}          & \multicolumn{4}{l|}{Noise generated by less meaningful annotations}                                                                                    & \multicolumn{1}{c|}{1}                  & \multicolumn{1}{c|}{0}                      & 1                                  \\
\multicolumn{1}{l|}{PUW-3}          & \multicolumn{4}{l|}{I can find test packages but cannot infer the coverage}                                                                            & \multicolumn{1}{c|}{2}                  & \multicolumn{1}{c|}{0}                      & 2                                  \\
\multicolumn{1}{l|}{PUW-4}          & \multicolumn{4}{l|}{Tool is more useful if the developer is familiar with schemas}                                                                     & \multicolumn{1}{c|}{4}                  & \multicolumn{1}{c|}{1}                      & 5   \\ \hline                             
\end{tabular}
\end{adjustbox}
\label{tab:coding}
\end{table}

\subsection{Program Comprehension}

\begin{figure}[h]
    \centering
    \includegraphics[width=\linewidth]{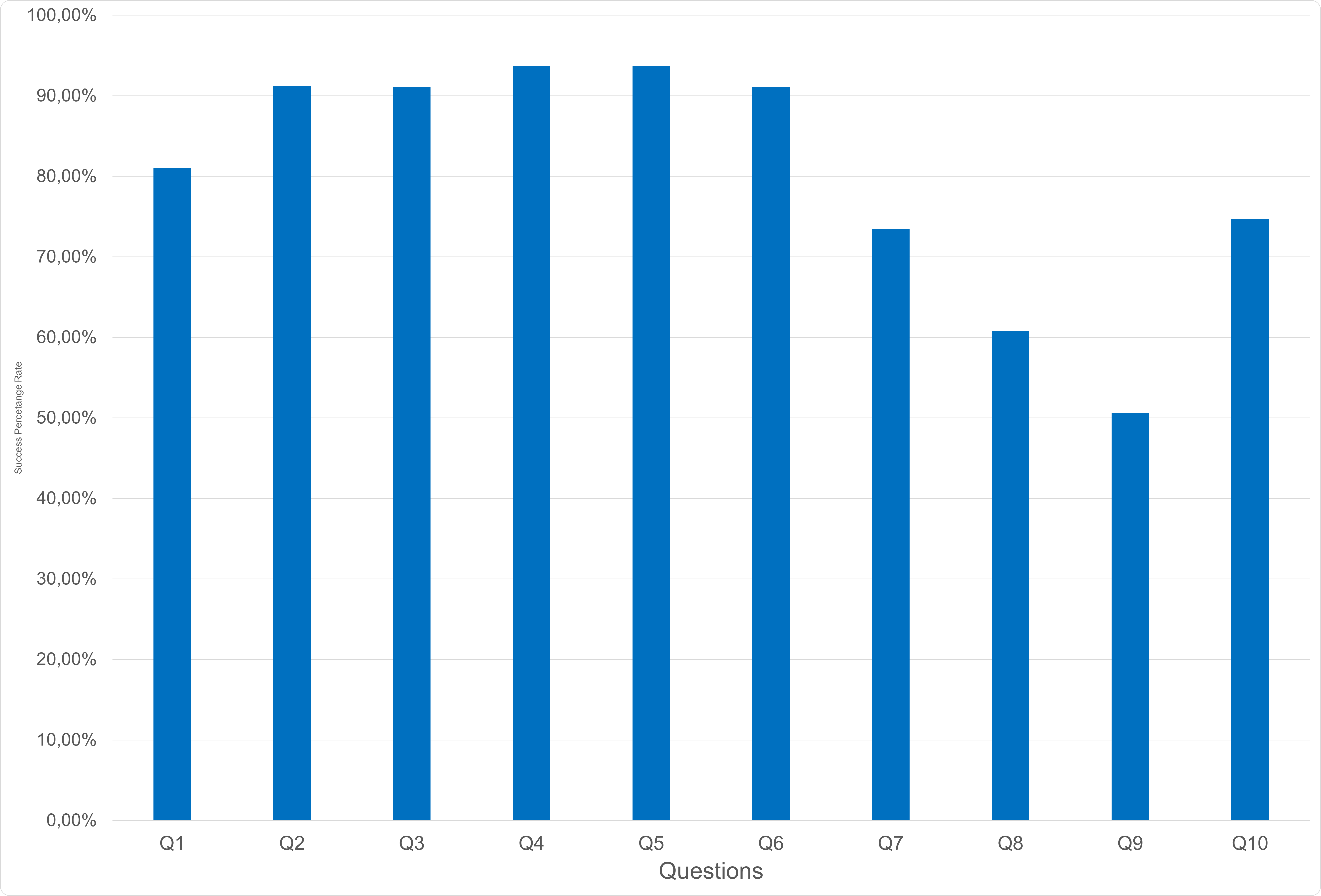}
    \caption{Program Comprehension Questionnaire}
    \label{fig:prog-comp-quest}
\end{figure}

The number of correct answers from the program comprehension questionnaire~\cite{zenodo} is graphically displayed in Figure \ref{fig:prog-comp-quest}. We used a bar chart to display the correctness of each question by all participants of the questionnaire. Questions Q2-Q6 had a high success rate, with roughly 90\% of the participants answering correctly. Questions Q1, Q6, and Q10 also performed well with 80\%, 75\%, and 73\% success rates, respectively.

Most of the questions with a high success rate are related to the \textbf{System View} and \textbf{Package View}. In other words, these are questions related to the general view of the system instead of a specific class or type. The \textcolor{black}{participants} could answer these without navigating far on the tool. The exception was Q4, which asked ``\textbf{What class contains the highest number of javax.persistence annotations?}'' which required the \textcolor{black}{participants} to reach the \textbf{Class View} and still performed very well with a 93\% of success rate. The coding presented in Table  \ref{tab:coding} can be used to sustain this result. \textcolor{black}{The theme PES-1 - \textbf{Easy to see annotation’s usage} and PUS-5 - \textbf{Detection of Responsibilities Using Schemas}, suggests that even in the \textbf{Class View}, the schema color information is still present, which helps detect the responsibilities. As seen in Table \ref{tab:coding}, PUS-5 had seven mentions, the third most present theme. This result helps confirm that there is a relationship between code annotations and class/package responsibilities as presented in Section~\ref{sec:background}. Finally, the CADV allows the visualization of such annotations, which helps identify and detect class/package responsibilities.} 

Questions Q8 and Q9, however, performed below the 70\% correctness rate. These questions required the user to investigate more deeply using the \textbf{Package View} and the \textbf{Class View}. Furthermore, they require a good understanding of the code annotation metrics instead of Q4, which asked about responsibilities. As found in Table \ref{tab:coding}, the theme PEW-3 appeared in 17 answers mentioning \textbf{Confusion when switching metrics between views}. Also, the theme PEW-5 - \textbf{Confusing how to trigger a new view} was mentioned five times. These results support the lower rate of correct answers for questions Q8 and Q9 that required changing views and understanding these metrics. The theme PEW-6 also helps explain these results because it implied that users were not checking the header to see what annotation metric was used.

Compared to Francese et al.~\cite{francese2016}, they had a similar correctness rate, ranging from 70\% to 87\%. However, we had higher values (93\%) in some questions, and one question performed lower (51\%).

\subsection{Perceived Ease of Use}

\begin{figure}[h]
    \centering
    \includegraphics[width=\linewidth]{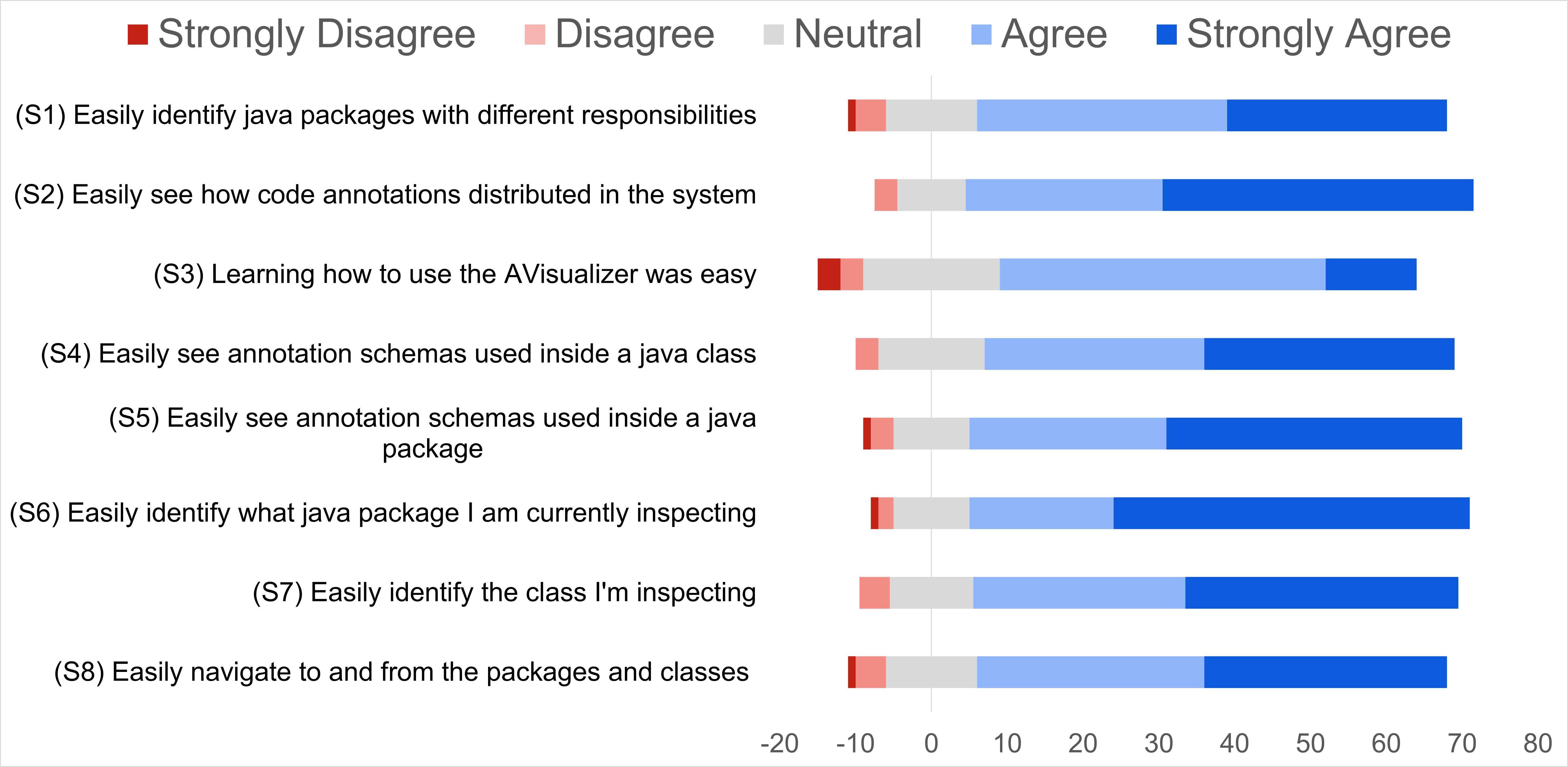}
    \caption{Perceived Ease of Use - Questionnaire}
    \label{fig:ease-of-use-quest}
\end{figure}

\textcolor{black}{To measure the perceived ease of use in the questionnaire with students, we issued eight statements presented in Table \ref{tab:tam-questionnaire} using the Likert Scale. Figure \ref{fig:ease-of-use-quest} displays a diverging bar chart with the results. Table \ref{tab:coding} presents the coding/themes that were in the majority extracted from the interviews with developers but also the one open-ended question students answered in the survey.}

From Figure~\ref{fig:ease-of-use-quest} the majority of the respondents \textbf{agree} or \textbf{strongly agree} that, overall, it is easy to use and obtain information from the target system. The statement S3 ``Learning how to use the AVisualizer was easy'' had the highest \textbf{strongly disagree} value, suggesting the tool requires training to be used appropriately. The theme PEW-7, which appeared in eight answers, states that training is required to use the tool and supports this result.

The statement S6 had the highest ``strongly agree'' value, suggesting it was easy localizing in what part of the software the user was. This may seem contradictory to the theme PEW-6, which suggests the headers were sub-utilized. However, crossing this data with the Program Comprehension results, themes PEW-5 and PEW-3, the header most likely failed to inform what metric was being used rather than what package or class the user was inspecting. Furthermore, PES-3 reinforces that the circle packing provided a good comprehension of the hierarchical structure, helping users locate themselves while using the tool.

In the strength theme, the most common theme in eleven answers was PES-2 claiming that it was \textbf{Easy to understand how to use the tool}. Then, in six answers, the respondents said it was \textbf{easy to see annotations' use} (PES-1). Three respondents also mentioned that the tool has a \textbf{quick learning curve} - PES-4.

In the weaknesses theme, we grouped codes considering aspects that make the tool harder to use. The most prominent was PEW-3 claiming it was \textbf{Confusing when switching metrics between views}, with 17 answers, followed by PEW-1 \textit{Understand the concept of annotation schema}. The code PEW-3 is also related to PEW-8 \textbf{Knowledge of Annotation Metrics}. The issue is that most users are not familiar with the code annotation metrics. PEW-1 suggests that even though code annotations are very popular among Java programmers, some may not know the concept of annotation schema. If a user is familiar with schemas, it becomes easier to use the tool. On the other hand, if the user lacks familiarity with schemas, it is harder to reason and extract information about them. 

\subsection{Perceived Usefulness}

\textcolor{black}{To measure the perceived usefulness from the students, we presented them with four scenarios, mentioned in Section \ref{sec:ease-use-usefulness-evaluation}. From the results, 41\% of the students would employ it in Scenario I, i.e., to detect schemas used in a given project and further learn about them. In second, with 34.62\%, was Scenario IV, ``Familiarize with the system before adding new features''. In third place, with 12.82\%, was Scenario III, ``Analyze the architecture''. The least chosen was Scenario II, related to searching for problems or misplaced annotations. These results suggest that the students perceive the tool. useful to support a learning process.}

Analyzing Table \ref{tab:coding}, the most common theme for strengths, identified in eleven answers, is PUS-1 \textbf{Visualization is better than code inspection to search annotations related information}, \textcolor{black}{suggesting the AVisualizer may outperform manual code inspection.} As a respondent wrote: \textit{``It is also a great idea since checking the code (going through dozens of packages) takes quite some time and is not visual, so this tool helps [a lot].''} Another interviewee mentioned that: \textit{``[...] when you put it visually, it is much easier for me to understand how they are spread in the system [...]''}. Besides that, the theme PUS-3 suggests that the tool can aid in refactoring and modularizing the system's architecture. Theme PUS-4 mentions that the AVisualizer can detect the simultaneous usage of frameworks with the same function. Even though these frameworks can usually work together without errors, using only one would make the solution more consistent across different features in the architecture. Finally, theme PUS-6 reinforces that the color strategy used to identify schemas can also help identify potentially misplaced code annotations. 

Detecting misplaced annotations may benefit developers when performing a refactoring process. As mentioned, the work of Yu~\cite{Yu2019} demonstrated that 19.8\% of annotation replacements are related to ``same name'' changes. In other words, the developers identified that the annotation name was correct. However, it belonged to an unexpected schema. With the AVisualizer, these scenarios can be spotted because an ``unexpected'' or ``different'' color circle will be in the wrong package. As stated by one interviewee: \textit{``[...] it is very simple to detect potentially misplaced code annotations. If I spot a pink circle in a package with only orange circles, I would say something is wrong. Why is a \texttt{javax.persistence} code annotation alone in a package with only \texttt{org.springframework} code annotations? '}'   

For the weakness, the most common code, present in 12 answers, was PUW-1 \textbf{I can find large annotations, but this is not very useful isolated}. These answers were rising when discussing topics such as finding large annotations or excessively used annotations. Even though the AVisualizer allows the detection of large annotations, the respondents did not see much value in this information isolated, claiming other design decisions outside the scope of the AVisualizer should also be used. This result differed from detecting misplaced annotations, which was seen as very useful.

One interesting code was PUW-2, suggesting that annotations such as \texttt{@Override} is trivial and probably pollute the visualization. Finally, present in the five answers was PUW-4: \textbf{tool is more useful if the developer is familiar with schemas}. As one respondent stated: \textit{``The colors help detect these potentially misplaced annotations. However, I argue that whoever is analyzing the system should also be familiar with code annotations and schemas. This way, a better decision is made to determine if the package or class requires further investigation or code inspection.''}

\subsection{Target Audience}

To validate the intended target audience, we asked the students what role they found the tool was suited for, allowing them to select multiple roles. The students chose ``Developer'' as the preferred role for using the AVisualizer, followed by ``Newcomer developer'' and ``Architect''. The least chosen was ``Project Manager''. Of the interviewees, three said ``Architect'' was most suited, while three said ``Every role in the team can benefit'', including the ``Architect''. 

One interviewee stated: \textit{``[...] the tool has different utilities within a software development team. The architect is usually concerned with the organization, and the tester may use it to detect packages that require more test code. A developer is usually more concerned with going straight to the code.''}. Another interviewee stated: \textit{``Everyone that touches code can benefit from this tool''}. However, one interviewee believes the newcomer should consider other aspects before, stating: \textit{``Newcomers should first study what design patterns are being used in the project. Verifying the annotations should be second''}.

\subsection{Visualization Goals Discussion}


\begin{itemize}

\item [(\#G1)]: To reach this goal, we designed the \textbf{System View}. The results show that both students and developers found this view the most useful and were able to get a general view of how code annotations were distributed in the system. 
    
    \item [(\#G2)]: To reach this goal, we designed the \textbf{Package View}. The results show similar results compared to the \textbf{System View}. Students and developers could visualize the usage and usage of code annotations within a specific package.
    
    \item [(\#G3)]: To reach this goal, we designed the \textbf{Class View}. The results show that this view did not reach the same correctness rate as the other two. In the qualitative analysis, we identified that this result was related to switching views and code annotation metrics unfamiliar to respondents. However, in the worst scenario, it reached correctness of 51\%.
    
    \item [(\#G4)]: The AVisualizer tool was developed with this feature, enabling users to navigate between all three views. From the results, the users could navigate between views after proper training. Although S8 presented a good agreement about how easy is the navigation in the tool, PEW-3 and PEW-5 showed that it could be improved.  
    
  \item [(\#G5)]: From the qualitative analysis, we found the tool is useful for detecting potentially misplaced annotations. As for detecting large or excessively used annotations, the coding revealed that the AVisualizer tool allows their detection, but respondents did not see much value in this. We argue that future experiments and a proper study should investigate code annotations bad smells and how these large annotations potentially impact software maintenance.

\end{itemize}

\subsection{Threats to Validity}

This section discusses the threats to the validity of our study. We followed the guidelines from Wohlin et al.~\cite{Wohlin2012} defining four steps to evaluate a study: conclusion validity, internal validity, construct validity, and external validity.

\subsubsection{Conclusion Validity}

The relationship between the treatment and the outcome, i.e., if the treatment is responsible for the outcome. A threat is that the respondents had five alternatives to the questions in the questionnaire, and some might have guessed the correct answer. For the interview, the developers were familiar with the underlying system and may have answered questions considering their previous knowledge of the system. We asked the interviewees to describe how they used the tool to obtain the answers to mitigate this. Finally, we conducted the study with two different groups of participants to mitigate potential bias from one group.

\subsubsection{Internal Validity}

In this step, the threat is identifying a causal relationship between two aspects when in reality, another aspect, not considered, is responsible for the consequence. In our study, an issue might be that the tasks were accomplished for other reasons than the tool.

All participants received the same training through video. However, the interviewees had a second live training session. Therefore they may have felt more comfortable than respondents to the questionnaire. During the evaluation, the participants had access to the video and could re-watch it to mitigate this threat. Furthermore, the questionnaire contained a guide explaining the annotation metrics.

\textcolor{black}{Two authors of this work were researchers that belonged to INPE. However, there was no direct relationship between them and the six interviewed developers of the SpaceWeatherTSI application, nor did the authors have any involvement in developing this application.}

\subsubsection{Construct Validity}

This step regards if we are correctly assessing the constructs under study. In our study, the key constructs are perceived ease of use and usefulness. To mitigate the threats of not correctly assessing them, we employed instruments developed for TAM that had been extensively used in research. The questionnaire was conducted remotely by the participants in their environment. They were also responsible for using the AVisualizer, watching the video tutorials, and answering the questions truthfully. Some participants might have had a more suitable environment or less interruption during the evaluation.

\subsubsection{External Validity}

This step regards generalization, i.e., if the results are valid beyond the subjects that participated in the study. Evaluations carried out with participants may have the ``evaluation apprehension'' threat, in which some might think they were being tested or evaluated and perform poorly. In order to mitigate that, especially with students, it was made clear that their answers to the questionnaire would not influence their grades. Another threat is that the chosen target software system displayed characteristics aligned with the goals of the AVisualizer. Not every Java software system is suitable to be visualized with the CADV. To mitigate any concerns, we clarified to participants that the AVisualizer was designed to visualize code annotations. Finally, a threat to the CADV approach is the heavy use of colors, severely impacting its usage by colorblind users. Currently, the approach revolves around colors to identify annotation schemas, being a core of the CADV approach. 
\section{Concluding Remarks}
\label{sec:conclusion}

We proposed a software visualization approach named Code Annotations Distribution Visualization (CADV) for code annotations. It is composed of three views based on circle packing representing the hierarchical structure of the system under analysis. The leaf nodes represent information about code annotations, and the size is calculated using a code annotation metric. The color strategy to represent annotation schemas allowed users to visualize how their responsibilities were distributed in the system under analysis.

As a reference implementation of the CADV, we developed a web-based tool called AVisualizer. To evaluate our visualization approach, we conducted an empirical study with students through a questionnaire and professional developers through an interview. The data was analyzed using both quantitative and qualitative approaches. We extracted the correctness of program comprehension tasks from the questionnaire, perceived ease of use, perceived usefulness, and evaluated the suitability for the intended target audience. The qualitative approach was carried out through a thematic analysis of the interview transcriptions and the questionnaire's open questions, which supported the results of the quantitative analysis. Our findings suggest that CADV is suitable for representing code annotations' distribution and organization, identifying responsibilities associated with schemas, and spotting potentially misplaced annotations. According to the participants, CADV is more appropriate for these tasks than the alternatives currently available to developers, which rely on code inspection aided by IDE code search features.

According to Merino et al.~\cite{merino2018}, it is not common for software visualization studies to conduct empirical evaluations. Our findings demonstrate more generally that software visualization approaches can aid in the software comprehension process. 

The current implementation of the AVisualizer supports only Java code annotations. However, it can be adapted to display C\# attributes by associating the schema with namespaces used to define C\# attributes. As for the circle packing approach is worth investigating how traditional source code metrics would be displayed in this kind of visualization. 

\textcolor{black}{Finally, the AVisualizer is a tool in constant evolution, and several new features are currently being added, mainly for customization purposes and to aid in identifying other potential annotation's bad practices. As of this point, a plugin for the IntelliJ IDE was developed\footnote{\url{https://github.com/metaisbeta/intelliJ-avisualizer-plugin}} so developers can use the AVisualizer embedded in an IDE. The demonstration can be downloaded directly from the Marketplace\footnote{\url{https://plugins.jetbrains.com/plugin/18237-annotation-visualizer}}. Furthermore, a feature to allow navigation from the tool to the source code is currently a work in progress.}

\section*{Acknowledgments}

We would like to thank the support granted by Brazilian funding agency FAPESP (São Paulo Research Foundation) grant 2019/12743-4

\bibliography{main}

\end{document}